\documentclass[12pt]{article}
\newcommand{\BE}{\begin{eqnarray}}
\newcommand{\EE}{\end{eqnarray}}
\newcommand{\be}{\begin{eqnarray}}
\newcommand{\ee}{\end{eqnarray}}
\usepackage{amssymb}
\usepackage{amsmath}
\usepackage{array}
\usepackage{latexsym}
\usepackage{epsfig}

\begin{document}
\vspace*{-.6in} \thispagestyle{empty}
\begin{flushright}
CALT-68-2523\\
UK-04-21
\end{flushright}
\baselineskip = 18pt

\vspace{1.0in} {\Large
\begin{center}
Gauge/Gravity Duality for Interactions of Spherical Membranes in 11-dimensional
pp-wave
\end{center}} \vspace{.5in}

\begin{center}
Hok Kong Lee$^1$, Tristan McLoughlin$^1$, and Xinkai Wu$^2$
\\
\emph{$^1$California Institute of Technology\\ Pasadena, CA  91125,
USA\\
$^2$Department of Physics and Astronomy\\
University of Kentucky, Lexington, KY  40506, USA\\
hok@theory.caltech.edu, tristanm@theory.caltech.edu,
xinkaiwu@pa.uky.edu}
\end{center}
\vspace{0.4in}

\begin{center}
\textbf{Abstract}
\end{center}
\begin{quotation}
\noindent We investigate the gauge/gravity duality in the
interaction between two spherical membranes in the 11-dimensional pp-wave
background. On the supergravity side, we find the solution to the
field equations at locations close to a spherical source membrane,
and use it to obtain the light cone Lagrangian of a spherical probe
membrane very close to the source, i.e., with their separation much
smaller than their radii. On the gauge theory side, using the BMN
matrix model, we compute the one-loop effective potential between two
membrane fuzzy spheres. Perfect agreement is found between the two
sides. Moreover, the one-loop effective potential we obtain on the
gauge theory side is valid beyond the small-separation
approximation, giving the full interpolation between interactions of
membrane-like objects and that of graviton-like objects.
\end{quotation}

\newpage

\pagenumbering{arabic}

\section{Introduction}

In this paper we will investigate the gauge/gravity duality in the
maximally supersymmetric eleven-dimensional pp-wave background. The
gauge theory side is represented by a 1d matrix theory first
proposed in \cite{Berenstein:2002jq} while the other side is
represented by 11d supergravity. Similar investigations have been
performed in flat space for various M-theory objects \cite{BB1, BB2,
BB3, BBPT, OY1, OY2, Okawa}. In particular, we will be interested in
the two-body interactions of point like gravitons and spherical M2
branes.

In pp-wave there is an interesting new connection we could make
between a graviton and an M2 brane. Under the influence of the
3-form background whose strength is proportional to a parameter
$\mu$, each stable M2 brane curls up into a sphere, with its radius
$r_0$ proportional to $\mu$ and its total momentum in the $x^-$
direction, i.e. we have $r_0 \sim \mu P_-$. If one takes the limit
of $\mu \rightarrow 0$ while keeping the total momentum $P_-$ fixed,
then the radius of the sphere goes to zero and we get a point like
graviton. If on the other hand, we increase $P_-$ simultaneously
such that the radius goes to infinity, then the end product will be
a flat membrane instead. Thus the gravitons and the flat membranes
of flat space could both be regarded as different limit of spherical
membranes in pp-wave. One could make another observation by considering
two spherical membranes separated by a distance $z$. If $z \gg r_0$,
then one expects the membranes to interact like two point-like
gravitons. If $z \ll r_0$ then the interaction should be akin to
that between flat membranes. Therefore by computing the interactions
between membranes of arbitrary radii and separation, we could then
take different limit to understand the interactions of both gravitons and
membranes.

In this paper we will compare the light-cone Lagrangian of
supergravity with the effective potential of matrix theory. With a
slight abuse in terminology, we will sometimes refer to both as the
effective potential. On the supergravity side, we will use linear
approximation when solving the field equations. This means all the
metric components computed this way will be proportional to no
higher than the first power of $\kappa_{11}^2$, higher order effects
such as recoiling and other back reactions could then be neglected.
On the matrix theory side, the interaction between two spherical
membranes begins at one loop, and for the purpose of comparing to
linearized supergravity, only the one-loop effective potential is
needed. A more detailed discussion can be found in
\cite{Lee:2003kf}.

Due to the complexity of the field equations of 11d supergravity, we
will only compute the effective potential of the supergravity side
in the near membrane limit ($z \ll r_0$). The graviton limit ($z \gg
r_0$) was already computed in our previous paper \cite{Lee:2003kf}.
On the matrix theory side, however, it is possible to compute the
expression for general $z$ and $r_0$, and by taking the appropriate
limits, we are able to find perfect agreement with supergravity. In
other words, the matrix theory result provides a smooth
interpolations between the near membrane limit and the graviton
limit. We  also compare our results with the those of Shin and
Yoshida \cite{Shin:2003np, Shin:2004az} and where there is overlap
we again have perfect agreement.

This paper is organized as follows. We begin in section
\ref{MTSugraIntro} with a brief discussion of the theories on both
sides of the duality. The relations between the parameters on the
two sides as well as a discussion on the special limits of interests
can be found here. In section \ref{EffP} we first define the probe
membrane's light cone Lagrangian on the supergravity side for a
general background, and then consider it in the pp-wave background
perturbed by a source. Next in section \ref{sec:sugra} the
linearized field equations are first diagonalized for an arbitrary
static source, and then solved in the special case of a spherical
membrane source in the near membrane limit ($z \ll r_0$). The metric
and the three-form potential are then used to compute the
supergravity light cone Lagrangian. Section \ref{gaugeside} is
devoted to computation on the matrix theory side. The one-loop
effective potential is found by integrating over all the fluctuating
fields. The near membrane limit of the resulting potential is
compared with the supergravity light cone Lagrangian and agreement
is found. In section \ref{Interpolation} we will compute the
one-loop interpolating effective potential on the matrix theory side
for arbitrary separation that takes us between the membrane limit
and the graviton limit. The result is compared with our earlier work
\cite{Lee:2003kf} as well as with that of Shin and Yoshida
\cite{Shin:2003np, Shin:2004az}. This is followed by a discussion in
section \ref{sec:discussion}.

\section{The Two Sides of the Duality} \label{MTSugraIntro}

\subsection{The Spherical Membranes}\label{subsection:sphericalmembrane}

The nonzero components of the maximally supersymmetric eleven-dimensional pp-wave metric and
the four-form field strength are given by:
\begin{equation}
g_{+-}=1,\ g_{++}
=-\mu^2\left[\frac{1}{9}\sum_{i=1}^3(x^i)^2+\frac{1}{36}
\sum_{a=4}^9(x^a)^2 \right],\ g_{AB}=\delta_{AB}
\end{equation}
\begin{equation}
F_{123+}=\mu
\end{equation}
The index convention throughout this paper is: $\mu,\nu,\rho,\ldots$
take the values $+,-,1,\ldots, 9$;  $A,B,C,\ldots$ take the values
$1,\ldots,9$; $i,j,k,\ldots$ take the values  $1,\ldots,3$; and
$a,b,c,\ldots$ take the values $4,\ldots,9$.

The matrix theory action in such a background is known
\cite{Berenstein:2002jq}:

\begin{eqnarray} \label{Maction}
{\cal S}=\int dt Tr\Bigg\{\sum_{A=1}^9\frac{1}{2R}(D_t X^A)^2+i
\psi^TD_t \psi + \frac{(M^3R)^2}{4R}\sum_{A,B=1}^9[X^A,X^B]^2  \nonumber\\
- (M^3R) \sum_{A=1}^9  \psi^T\gamma^A[\psi,X^A] +
\frac{1}{2R}\left[-(\frac{\mu}{3})^2\sum_{i=1}^3(X^i)^2
-(\frac{\mu}{6})^2\sum_{a=4}^9(X^a)^2\right]  \nonumber\\
-i \frac{\mu}{4}\psi^T\gamma_{123}\psi
-\frac{(M^3R)\mu}{3 R}i\sum_{i,j,k=1}^3\epsilon_{ijk} X^iX^jX^k
\Bigg\}
\end{eqnarray}
where $D_t X=\partial_tX^I-i[X_0,X^I]$. The constant $M$ in the
action above is the eleven-dimensional Planck mass, and $R$ is the
compactification radius in the $X^-$ light-like direction in the
DLCQ formalism.

The supersymmetric configurations of this action have been well studied,
see for example \cite{Park:2002cb}. Although in the end we will add
a perturbation in the $X^4$ to $X^9$ directions to make it
non-supersymmetric, we begin with the following configuration:
\begin{eqnarray}
X_i &=& \frac{\mu}{3 M R^3} J_i \nonumber \\
X_a &=& 0
\end{eqnarray}
where $[J_i, J_j] = i \epsilon_{ijk} J_k$.

If $J_i$ is an $N\times N$ irreducible matrix, then it represents a single sphere with radius $r_0$ given by:
\begin{eqnarray}\label{eqn:r0_matrix}
r_0 = \sqrt{\frac{1}{N} Tr X_i^2} = \frac{\mu}{6 M^3 R} \sqrt{N^2
-1} \approx \frac{\mu N}{6 M^3 R}
\end{eqnarray}
The last approximation is taken with $N$ large, such that the matrix
theory correction to the radius (denoted as $\delta r_0$) is negligible
compared to $r_0$. In other words, we choose $N$ large
enough that $\frac{\delta r_0 }{r_0} \sim \frac{1}{N^2}
\rightarrow 0$. The reader is reminded that the purpose of this
paper is to compare matrix theory to the predictions of
supergravity, so we are not interested in any finite $N$ effects
which are related to matrix theory corrections to supergravity.

If $J_i$ is reducible it represents multiple concentric
spherical membranes, and the radius of each irreducible component can
be found in the same way as for the case above.

The configuration we are going to use is one where $J_i$ is the
direct sum of two irreducible components. This represents two
membranes of different radius. In section \ref{gaugeside} we will
put in non-trivial $X_a$ to break the supersymmetry, which
physically corresponds to placing one of the membranes away from the
origin.

On the supergravity side, the bosonic part of the Lagrangian of a
probe membrane in a general background is  \BE{\cal
L}(X^\mu,\partial_iX^\mu)=-T\left[\sqrt{-\det(g_{ij})}
-\frac{1}{6}\epsilon^{ijk}A_{\mu\nu\rho}\partial_iX^\mu\partial_jX^\nu\partial_kX^\rho\right]\EE
with $T$ being the membrane tension, and $g_{ij}\equiv
G_{\mu\nu}\partial_iX^\mu\partial_jX^\nu$ being the pullback metric.

One can define the light cone Lagrangian ${\cal L}_{lc}$ via a
Legendre transformation in the $X^-$ direction (see Section
\ref{EffP}). In the unperturbed pp-wave background, we have
\begin{eqnarray}\label{eqn_Llcpp}({\cal L}_{lc})_{pp}&
&=\frac{1}{2}\Pi_-\partial_0X^A\partial_0X^A
-\frac{\Pi_-}{2}\mu^2\left[\frac{1}{9}(X^i)^2+\frac{1}{36}(X^a)^2\right]\nonumber\\
&
&-\frac{T^2}{4\Pi_-}(\partial_1X^A\partial_2X^B-\partial_1X^B\partial_2X^A)^2
+T\frac{\mu}{3}\epsilon_{ijk}(\partial_1X^i)(\partial_2X^j)X^k\end{eqnarray}
with $\Pi_-$ being the momentum density in the $X^-$ direction.
Looking at $({\cal L}_{lc})_{pp}$, above, wee see that one solution
to the equations of motion in the unperturbed pp-wave is a spherical
membrane at rest with $X^iX^i=r_0^2$, $X^a=0$, $X^+=t$,
$X^-=0$, with its radius being $r_0=\frac{\mu\Pi_-}{3T \sin\theta}$
(note that we take the worldvolume coordinates $\sigma^{0,1,2}$ to
be $t,\theta,\phi$; also recall that it is
$\frac{\Pi_-}{\sin\theta}$ that is a constant on the worldvolume).
In terms of the total momentum $P_-=\int d\theta d\phi \Pi_-$, this
radius is \BE r_0=\frac{\mu P_-}{12\pi T}\EE

%


Later in this paper, we will take the background to be the pp-wave
perturbed by a source membrane:
$G_{\mu\nu}=(G_{\mu\nu})_{pp}+h_{\mu\nu}$, and
$A_{\mu\nu\rho}=(A_{\mu\nu\rho})_{pp}+a_{\mu\nu\rho}$. Then ${\cal
L}_{lc}=({\cal L}_{lc})_{pp}+\delta {\cal L}_{lc}$, and we shall
compare $\delta {\cal L}_{lc}$ with the one-loop effective potential
on the matrix theory side (while (${\cal L}_{lc})_{pp}$ agrees with
the tree level Lagrangian on the matrix theory side, of course).

Identifying the radius calculated on the two sides and $P_-$ with
$N/R$, we get the following relation between the tension of the
membrane and the Planck mass:
\begin{eqnarray}
2 \pi T = M^3
\end{eqnarray}

Another relation we will use to compare the two side is the
identification $\kappa_{11}^2 = \frac{16 \pi^5}{M^9}$
\cite{Lee:2003kf, Becker:1997xw}, and for convenience we will define
the parameter $\alpha = \frac{1}{M^3 R}$ that will appear often on
the matrix theory side.

\subsection{The Effective Potential}\label{subsection:form_of_V}
As already stated, on the supergravity side the quantity of interest
is the probe light cone Lagrangian. On the matrix theory side, the
one-loop effective potential is the relevant quantity, and is
defined through the Euclideanized effective action with all
quadratic fluctuations integrated out. In the following, the
subscripts $_M$ and $_E$ will be used to denote Minkowski and
Euclidean signature respectively.

Euclideanization is carried out by defining the following Euclideanized quantities:
\begin{eqnarray}
\tau_E&=&i\tau_M \nonumber \\
{\cal S}_E &=& -i {\cal S}_M
\end{eqnarray}

The Euclideanized action ${\cal S}_E $ will then be expanded about a certain background (to be specified in section \ref{gaugeside}) and the quadratic fluctuations integrated out to produce the one-loop effective action $\Gamma_E$. The effective action on the matrix theory side is then defined via the relation:
\begin{eqnarray}
\Gamma_E = -\int d\tau_E V_{eff}
\end{eqnarray}

The minus sign in front of the integral is slightly unconventional,
but it was put there for the convenience of comparison with
supergravity. It was chosen such that the tree level part of the
effective potential is simply the light cone Lagrangian $({\cal
L}_{lc})_{pp}$ rather than $-({\cal L}_{lc})_{pp}$. After $V_{eff}$
is computed, the result could then be analytically continued back
into Minkowski signature by replacing $v_E \rightarrow i v_M$.

In order to facilitate the comparison of the two sides, it is
useful to first examine the form of the action. The supergravity effective action is given in eqn(\ref{eqn:deltaLlc_result}). For the simple case when the two membranes have the same radius and in the limit where their separation in the $X^4$ to $X^9$ direction, $z$, is small, i.e. $z \ll r_0$, the supergravity effective action is given in eqn(\ref{eqn:deltaLlc_result}). Putting $w$, the difference of the two radii to zero, and not keeping track of the exact coefficients, we could rewrite the supergravity result in term of matrix theory parameters:
\begin{eqnarray}
V_{eff}= \alpha (\frac{v^4}{\mu ^2 z^5}+\frac{v^2}{
z^3}+\frac{\mu^2}{z} )
\end{eqnarray}
Hence we see that a comparison with supergravity means looking at
order $(\alpha)^1$ on the matrix theory side.

\subsection{The Membrane Limit and the Graviton Limit}

We will first state the two limits we are interested in:\\

Membrane limit:
\begin{eqnarray}
\frac{z}{\alpha \mu} &\gg& 1 \\ \label{ML1}
\frac{z}{\alpha \mu} &\ll& N
\end{eqnarray}

Graviton limit:
\begin{eqnarray}
\frac{z}{\alpha \mu} &\gg& 1 \\
\frac{z}{\alpha \mu}&\gg& N
\end{eqnarray}
where we used $z$ to denote the separation of the two spherical
membranes in the $X^4$ to $X^9$ directions, and recall $\alpha=
\frac{1}{M^3 R}$.

The membrane limit is derived from the condition:
\begin{eqnarray}
\frac{1}{N} \ll \frac{z}{r_0} \ll 1
\end{eqnarray}
The first inequality ensures that the effect of non-zero $z$ is
greater than any matrix theory corrections to supergravity, which we
are not interested in. The second inequality ensures we are at the
near membrane limit. Using $r_0=\frac{\alpha\mu N}{6}$ (see eqn
(\ref{eqn:r0_matrix})), we arrive at the limit as stated.

The graviton limit is when $z$ is much greater than $r_0$, so that
approximately the two spheres interact like two point like
gravitons. We still enforce the condition $\frac{1}{N} \ll
\frac{z}{r_0}$ for comparison with supergravity but reverse the
second inequality in the membrane limit to $\frac{z}{r_0} \gg 1$ to
arrive at the graviton limit stated above.

In this paper, on the supergravity side we will calculate the light
cone Lagrangian only in the membrane limit. The light cone
Lagrangian in the graviton limit was already computed in our earlier
work \cite{Lee:2003kf}. On the matrix theory side, the one-loop
effective potential could be computed for general $z$. The two
limits of this potential is then compared with the supergravity side
and we will find perfect agreement. Later in the paper we will use
matrix theory to find explicitly the potential that interpolates
between the two limits.

\section{The Supergravity Light Cone Lagrangian}\label{EffP}

The light cone Lagrangian ${\cal L}_{lc}$ of the probe is basically
its Lagrangian Legendre transformed in the $x^-$ degree of freedom.
Here we briefly derive the ${\cal L}_{lc}$ for a probe membrane in a
pp-wave background perturbed by some source. (The reader is referred
to Section 4.1 of \cite{xinkaithesis} for the detailed discussion of
the light-cone Lagrangian of point particle probes and membrane
probes in terms of Hamiltonian systems with constraints.)

Denote the background metric and three-form as $G_{\mu\nu}(x)$,
$A_{\mu\nu\rho}(x)$, respectively, and the membrane embedding
coordinates as $X^\mu(\sigma^i)$, with $\sigma^i,i=0,1,2$ being the
world-volume coordinates. The bosonic part of the membrane
Lagrangian density is given by \BE{\cal
L}(X^\mu,\partial_iX^\mu)=-T\left[\sqrt{-\det(g_{ij})}
-\frac{1}{6}\epsilon^{ijk}A_{\mu\nu\rho}\partial_iX^\mu\partial_jX^\nu\partial_kX^\rho\right]\EE
with $T$ being the membrane tension, and $g_{ij}\equiv
G_{\mu\nu}\partial_iX^\mu\partial_jX^\nu$ being the pullback metric.
The momentum density is \BE\Pi_\lambda\equiv\frac{\partial {\cal
 L}}{\partial (\partial_0X^\lambda)}
 =- T\left[\sqrt{-\det(g_{ij})}\ g^{0k}(\partial_kX^\mu)G_{\lambda\mu}
 -A_{\lambda\nu\rho}\partial_1X^\nu\partial_2X^\rho \right]\EE
Define \BE\tilde{\Pi}_\lambda\equiv\Pi_\lambda
-TA_{\lambda\nu\rho}\partial_1X^\nu\partial_2X^\rho
=-T\sqrt{-\det(g_{ij})}\ g^{0k}(\partial_kX^\mu)G_{\lambda\mu}\EE
There are the following three primary constraints \BE &
&\phi_0\equiv G^{\lambda\xi}\tilde{\Pi}_\lambda\tilde{\Pi}_\xi
+T^2\det(g_{rs})=0\nonumber\\
& &\phi_r\equiv\Pi_\lambda\partial_rX^\lambda\EE where $r,s=1,2$
label the spatial world-volume coordinates.

Now use the light-cone coordinates $\{x^+,x^-,x^A\}$, and assume
static, i.e., $x^+$-independent, background $G_{\mu\nu}(x^-,x^A)$
and $A_{\mu\nu\rho}(x^-,x^A)$. Use the light cone gauge
$X^+=\sigma^0$. Then the equation of motion for $X^+$ fixes the
Lagrange multiplier $c^0$ for the constraint $\phi_0$ \BE
c^0=\frac{1}{2G^{+\xi}\tilde{\Pi}_\xi}\EE The constraint $\phi_0=0$
(in which $X^+$ is set to $\sigma^0$) can be used to solve for
$\Pi_+$ as a function of
$(X^-,X^A,\partial_rX^-,\partial_rX^A,\Pi_-,\Pi_A)$.

We shall also express $\Pi_A$ as a function of
$(X^-,X^A,\partial_rX^-,\partial_rX^A,\Pi_-,\partial_0X^A)$ by
inverting the equations of motion for $X^A$: \BE\label{eqn:d0XA}
\frac{\partial X^A}{\partial
\sigma^0}=\frac{G^{A\xi}\tilde{\Pi}_\xi}{G^{+\nu}\tilde{\Pi}_\nu}
+c^r\frac{\partial X^A}{\partial\sigma^r}\EE where we have used the
expression for $c^0$ given above and the $c^r$'s are the Lagrange
multipliers for the $\phi^r$'s.

Then we can first define the light-cone Hamiltonian ${\cal
H}_{lc}(X^-,X^A,\partial_rX^-,\partial_rX^A,\Pi_-,\Pi_A)\equiv
-\Pi_+$, and then
 define the
light-cone Lagrangian \BE{\cal
L}_{lc}(X^-,X^A,\partial_rX^-,\partial_rX^A,\Pi_-,\partial_0X^A)
\equiv\Pi_A\partial_0X^A-{\cal H}_{lc}=\Pi_A\partial_0X^A+\Pi_+\EE

Now the background in which the probe membrane moves is the pp-wave
perturbed by some source: $G_{\mu\nu}=(G_{\mu\nu})_{pp}+h_{\mu\nu}$,
and $A_{\mu\nu\rho}=(A_{\mu\nu\rho})_{pp}+a_{\mu\nu\rho}$, where the
quantities with subscript $pp$ are those of the unperturbed pp-wave
background, and $h_{\mu\nu}, a_{\mu\nu\rho}$ are the perturbations
caused by the source. We only need the light-cone Lagrangian to
linear order in the perturbation. Solving $\phi_0=0$ gives \BE
\Pi_+(X^-,X^A;\partial_rX^-,\partial_rX^A;\Pi_-,\Pi_A)=(\Pi_+)_{pp}+\delta
\Pi_+\EE where \BE (\Pi_+)_{pp}=&
&\frac{-1}{2\Pi_-}\left\{-g_{++}\Pi_-^2+\Pi_A\Pi_A+\frac{T^2}{2}(
\partial_1X^A\partial_2X^B-\partial_1X^B\partial_2X^A)^2 \right\}\nonumber\\
& &+T\frac{\mu}{3}\epsilon_{ijk}(\partial_1X^i)(\partial_2X^j)X^k\EE
and \BE\label{eqn:deltaPi+} \delta\Pi_+=\frac{-1}{2\Pi_-}\Bigg\{&
&-h_{--}(\tilde{\Pi}_+)_{pp}^2
-2\Pi_-(h_{+-}-g_{++}h_{--})(\tilde{\Pi}_+)_{pp}-2T(\tilde{\Pi}_+)_{pp}a_{-\nu\rho}\partial_1X^\nu\partial_2X^\rho
\nonumber\\
& &-2T\Pi_- a_{+\nu\rho}\partial_1X^\nu\partial_2X^\rho-2h_{-
A}\Pi_A
(\tilde{\Pi}_+)_{pp}-\Pi_-^2[h_{++}+(g_{++})^2h_{--}-2g_{++}h_{+-}]\nonumber\\
& &+2T\Pi_-g_{++}a_{-\nu\rho}\partial_1X^\nu\partial_2X^\rho
-h_{AB}\Pi_A\Pi_B-2\Pi_-(h_{+A}-g_{++}h_{-A})\Pi_A\nonumber\\
& &-2T\Pi_A
a_{A\nu\rho}\partial_1X^\nu\partial_2X^\rho+T^2\delta\det(g_{rs})
\Bigg\}\nonumber\\
\EE where \BE
(\tilde{\Pi}_+)_{pp}=\frac{-1}{2\Pi_-}\left\{-g_{++}\Pi_-^2+\Pi_A\Pi_A+\frac{T^2}{2}(
\partial_1X^A\partial_2X^B-\partial_1X^B\partial_2X^A)^2 \right\}\EE

 The Lagrange multipliers $c^r$'s  are \BE c^r=(c^r)_{pp}+\delta
c^r=\delta c^r\EE where we have taken $(c^r)_{pp}$ to be zero.
Inverting eqn (\ref{eqn:d0XA}) gives \BE
\Pi_A(X^-,X^A;\partial_rX^-,\partial_rX^A;\Pi_-,\partial_0X^A)=(\Pi_A)_{pp}+\delta\Pi_A\EE
where $(\Pi_A)_{pp}=\Pi_-\partial_0X^A$, and $\delta \Pi_A$ depends
on $\delta c^r$.

Now \BE{\cal L}_{lc}=(\Pi_A)_{pp}\partial_0X^A+\delta
\Pi_A\partial_0X^A+(\Pi_+)_{pp}+\delta \Pi_+ \EE However, as can be
seen easily, the second term in the above expression is cancelled by
the part containing $\delta\Pi_A$ in the third term. So ${\cal
L}_{lc}$ does not contain $\delta\Pi_A$, thus being independent of
the $\delta c^r$'s.

Hence one finds \BE{\cal L}_{lc}=({\cal L}_{lc})_{pp}+\delta {\cal
L}_{lc}\EE with
\begin{eqnarray}\label{eqn_Llcpp}({\cal L}_{lc})_{pp}&
&=\frac{1}{2}\Pi_-\partial_0X^A\partial_0X^A
-\frac{\Pi_-}{2}\mu^2\left[\frac{1}{9}(X^i)^2+\frac{1}{36}(X^a)^2\right]\nonumber\\
&
&-\frac{T^2}{4\Pi_-}(\partial_1X^A\partial_2X^B-\partial_1X^B\partial_2X^A)^2
+T\frac{\mu}{3}\epsilon_{ijk}(\partial_1X^i)(\partial_2X^j)X^k\end{eqnarray}
and $\delta {\cal L}_{lc}=\delta\Pi_+$, where $\delta\Pi_+$ is given
by eqn (\ref{eqn:deltaPi+}) with all the $\Pi_A$'s in it replaced
by $(\Pi_A)_{pp}=\Pi_-\partial_0X^A$ (the reason being that
$\delta\Pi_+$ contains $h_{\mu\nu}$ and $a_{\mu\nu\rho}$ and is thus
already first-order). We'd like to emphasize again that ${\cal
L}_{lc}$ is independent of the choice of the $\delta c^r$'s.

To use the general expression for ${\cal L}_{lc}$ given in the
previous paragraph in a specific problem, one just have to substitute in
the $h_{\mu\nu}$'s and $a_{\mu\nu\rho}$'s found by solving the
supergravity field equations for the specific source, as well as the
specific probe membrane embedding $X^A(\sigma^i)$.

\section{Supergravity Computation}\label{sec:sugra}
\subsection{Diagonalizing the Field Equations for Arbitrary Static
Source}\label{subsec:diagonalsugra} In this subsection we present
the diagonalization of the linearized supergravity equations of
motions for arbitrary static sources (this subsection is basically
Section 4.2 of \cite{xinkaithesis}). There is, of course, no
highbrow knowledge involved here: we are just solving the linearized
Einstein equations and Maxwell equations, which are coupled; and by
``diagonalization'' we basically just mean the prescription by
which we get a decoupled Laplace equation for each component of the
metric and three-form perturbations. The unperturbed background is
the 11-D pp-wave, and we only consider static, i.e.,
$x^+$-independent, field configurations, thanks to the fact that the
sources considered are taken to be static, i.e., with
$x^+$-independent stress tensor and three-form current.

Since we leave the source arbitrary, what we'll present here are the
left-hand side of the linearized equations. These are tensors whose
computation is straightforward though a bit tedious: the reason we
present them here is because they are necessary when solving the
field equations, and to the best of our knowledge have not been
explicitly given elsewhere.

A somewhat related problem is the diagonalization of the equations
of motion when the source is absent. This requires field
configurations with $x^+$-dependence. One work along this line is
\cite{Kimura:2003um}. Roughly speaking, borrowing the language of
electromagnetism, what's considered in \cite{Kimura:2003um} are
electromagnetic waves in vacuum, while what we are considering here
are electrostatics and magnetostatics for arbitrary static sources.

The nonzero components up to (anti)symmetry of the Christoffel
symbol, Riemann tensor, and Ricci tensor of the 11-D pp-wave are \BE
& &\Gamma^{A}_{++}=-\ \frac{1}{2}\partial_A g_{++},\ \ \
\Gamma^{-}_{+A}=\frac{1}{2}\partial_A g_{++}\nonumber\\
& &R_{+A+B}=-\ \frac{1}{2}\partial_A\partial_B g_{++},\ \ \
R_{++}=-\ \frac{1}{2}\partial_C\partial_C g_{++}\EE (We usually
do not substitute the explicit expression of $g_{++}$, unless that
brings significant simplification to the resulting formula)

Now let's perturb the pp-wave background by adding a source. Denote
the metric perturbation $\delta g_{\mu\nu}$ by $h_{\mu\nu}$, and the
gauge potential perturbation by $\delta
A_{\mu\nu\rho}=a_{\mu\nu\rho}$. $h_{\mu\nu},a_{\mu\nu\rho}$ are
treated as rank-two and rank-three tensors, respectively, the
covariant derivative $\nabla$ acting on them is defined using the
connection coefficient of the unperturbed pp-wave background, and
indices are raised/lowered, traces taken using the background metric
$g_{\mu\nu}$.

Let's deal with the Einstein equations first. Define
$\bar{h}_{\mu\nu}\equiv h_{\mu\nu}-\frac{1}{2}g_{\mu\nu}h$, where
$h\equiv g^{\mu\nu}h_{\mu\nu}$. Without the source, the Einstein
equation is \BE
R_{\mu\nu}-\frac{1}{2}Rg_{\mu\nu}-\kappa_{11}^2[T_{\mu\nu}]_A=0\EE
Recall that the stress tensor of the gauge field is
\begin{equation} \left[T_{\mu\nu}\right]_A
=\frac{1}{12\kappa_{11}^2}\left( F_{\mu \lambda\xi\rho}F_\nu^{\
\lambda\xi\rho}-\frac{1}{8}g_{\mu\nu}F^{\rho\sigma\lambda\xi}F_{\rho\sigma\lambda\xi}\right)
\end{equation}
The source perturbs the Einstein equation to \BE
\delta\left(R_{\mu\nu}-\frac{1}{2}Rg_{\mu\nu}\right)-
\kappa_{11}^2\delta [T_{\mu\nu}]_A=\kappa_{11}^2[T_{\mu\nu}]_S\EE
with $[T_{\mu\nu}]_S$ standing for the stress tensor of the source.

As usual, it helps to proceed in an organized manner, grouping
different terms in the above perturbed Einstein equations. One
finds,
$\delta\left(R_{\mu\rho}-\frac{1}{2}Rg_{\mu\rho}\right)=-\frac{1}{2}\nabla^\sigma\nabla_\sigma
\bar{h}_{\mu\rho}+K_{\mu\rho}+Q_{\mu\rho}$, and $\kappa_{11}^2\delta
[T_{\mu\nu}]_A=N_{\mu\nu}+L_{\mu\nu}$, where the explicit
expressions of the symmetric tensors $\nabla^\sigma\nabla_\sigma
\bar{h}_{\mu\nu},K_{\mu\nu},Q_{\mu\nu}$,$N_{\mu\nu}$, and
$L_{\mu\nu}$ can be obtained after some work. Their definitions and
components are given below \footnote{Notice that $\partial_+$ will
never appear because we only consider the static case; also note
$g^{\mu\nu}\partial_\mu\partial_\nu=-g_{++}\partial_-^2+\partial_A\partial_A$
for static configurations.}
\newline
\newline
$\bullet \ \ \nabla^\sigma\nabla_\sigma \bar{h}_{\mu\nu}$ \BE
\nabla^\sigma\nabla_\sigma \bar{h}_{++}=&
&g^{\mu\nu}\partial_\mu\partial_\nu\bar{h}_{++}+\left[-(\partial_A\partial_A
g_{++})\bar{h}_{+-}+\frac{1}{2}(\partial_A g_{++}\partial_A
g_{++})\bar{h}_{--} \right]\nonumber\\ & &+2\left[\partial_A
g_{++}\partial_-\bar{h}_{+A}-\partial_A
g_{++}\partial_A\bar{h}_{+-}\right]\EE \BE
\nabla^\sigma\nabla_\sigma
\bar{h}_{+-}=g^{\mu\nu}\partial_\mu\partial_\nu\bar{h}_{+-}-\frac{1}{2}(\partial_A\partial_A
g_{++})\bar{h}_{--}+\partial_A
g_{++}\partial_-\bar{h}_{-A}-\partial_A
g_{++}\partial_A\bar{h}_{--}\nonumber\\
\EE \begin{equation} \nabla^\sigma\nabla_\sigma
\bar{h}_{+C}=g^{\mu\nu}\partial_\mu\partial_\nu\bar{h}_{+C}-\frac{1}{2}(\partial_A\partial_A
g_{++})\bar{h}_{-C}+\partial_A
g_{++}\partial_-\bar{h}_{AC}-\partial_C
g_{++}\partial_-\bar{h}_{+-}-\partial_A g_{++}\partial_A\bar{h}_{-C}
\end{equation} \BE \nabla^\sigma\nabla_\sigma
\bar{h}_{--}=g^{\mu\nu}\partial_\mu\partial_\nu\bar{h}_{--}\nonumber\\
\EE \BE \nabla^\sigma\nabla_\sigma
\bar{h}_{-C}=g^{\mu\nu}\partial_\mu\partial_\nu\bar{h}_{-C}-\partial_C
g_{++}\partial_-\bar{h}_{--} \EE \BE \nabla^\sigma\nabla_\sigma
\bar{h}_{CD}=g^{\mu\nu}\partial_\mu\partial_\nu\bar{h}_{CD}-\partial_C
g_{++}\partial_-\bar{h}_{-D}-\partial_D
g_{++}\partial_-\bar{h}_{-C}\EE

$\bullet \ \ K_{\mu\nu}$ \ \ Its definition is \BE
K_{\mu\rho}\equiv\frac{1}{2}\left(R_\mu^{\
\xi}\bar{h}_{\xi\rho}+R_\rho^{\
\xi}\bar{h}_{\xi\mu}\right)+R^{\sigma\ \ \ \xi}_{\
\mu\rho}\bar{h}_{\sigma\xi}+\frac{1}{2}g_{\mu\rho}R^{\xi\sigma}\bar{h}_{\xi\sigma}
-\frac{1}{2}R\bar{h}_{\mu\rho}\EE Its components are given by \BE
K_{++}=\left(-\frac{1}{2}\partial_A\partial_A
g_{++}\right)\left(\bar{h}_{+-}+\frac{1}{2}g_{++}\bar{h}_{--}\right)+\frac{1}{2}\left(\partial_A\partial_B
g_{++}\right)\bar{h}_{AB}\EE \BE
K_{+-}=\left(-\frac{1}{2}\partial_A\partial_A
g_{++}\right)\bar{h}_{--}\EE \BE
K_{+A}=\left(-\frac{1}{4}\partial_C\partial_C
g_{++}\right)\bar{h}_{-A}+\left(-\frac{1}{2}\partial_A\partial_B
g_{++}\right)\bar{h}_{-B} \EE \BE K_{--}=0\EE \BE K_{-A}=0\EE \BE
K_{AB}=\frac{1}{2}\left[\partial_A\partial_B
g_{++}-\frac{1}{2}\delta_{AB}\partial_C\partial_C g_{++}
\right]\bar{h}_{--}\EE

$\bullet \ \ Q_{\mu\nu}$ \ \ Its definition is
$Q_{\mu\rho}\equiv\frac{1}{2}(\nabla_{\mu}q_{\rho}+\nabla_{\rho}q_{\mu})-\frac{1}{2}g_{\mu\rho}\nabla^\alpha
q_{\alpha}$, where $q_{\alpha}\equiv
\nabla^\beta\bar{h}_{\beta\alpha}$. As one can recognize,
$Q_{\mu\rho}$ contains the arbitrariness of making different gauge
choices when solving the Einstein equation, where one makes a gauge
choice by specifying the $q_{\mu}$'s. The components of
$Q_{\mu\rho}$ are \BE& & Q_{--}=\partial_- q_-,\ \
Q_{-A}=\frac{1}{2}(\partial_- q_A+\partial_A q_-),\ \
Q_{-+}=\frac{1}{2}(g_{++}\partial_- q_- -\partial_A q_A)\nonumber\\
& & Q_{AB}=\frac{1}{2}(\partial_A q_B+\partial_B
q_A)-\frac{1}{2}\delta_{AB}(\partial_- q_+ -g_{++}\partial_- q_-
+\partial_A q_A)\nonumber\\ & &Q_{+A}=\frac{1}{2}\left[\partial_A
q_+ -(\partial_A g_{++})q_-\right]\nonumber\\&
&Q_{++}=\frac{1}{2}(\partial_A
g_{++})q_A-\frac{1}{2}g_{++}(\partial_- q_+ -g_{++}\partial_- q_-
+\partial_A q_A)\EE

$\bullet \ \ N_{\mu\nu}$ \ \ It is defined to be the part of
$\kappa_{11}^2\delta [T_{\mu\nu}]_A$ that contains only the metric
perturbation, but not the three-form gauge potential perturbation.
Its components are given by \BE & &
N_{++}=\mu^2\left(\frac{1}{3}\bar{h}_{+-}+\frac{1}{12}g_{++}\bar{h}_{--}-\frac{1}{3}\sum_{i=1}^3
\bar{h}_{ii}+\frac{1}{6}\sum_{a=4}^9\bar{h}_{aa}\right)\nonumber\\&
& N_{+-}=\frac{\mu^2}{4}\bar{h}_{--},\ \
N_{+i}=\frac{\mu^2}{2}\bar{h}_{-i}, \ \ N_{+b}=0\nonumber\\&
&N_{--}=0,\ \ N_{-i}=0,\ \ N_{-b}=0\nonumber\\&
&N_{ij}=-\frac{\mu^2}{4}\delta_{ij}\bar{h}_{--},\ \ N_{ib}=0,\ \
N_{ab}=\frac{\mu^2}{4}\delta_{ab}\bar{h}_{--}\EE

$\bullet \ \ L_{\mu\nu}$ \ \ This is defined to be the part of
$\kappa_{11}^2\delta [T_{\mu\nu}]_A$ that contains only the
three-form perturbation, but not the metric perturbation. Its
components are given by \BE& &L_{++}=\mu\left(\delta
F_{123+}-\frac{1}{2}g_{++}\delta F_{123-}\right),\ L_{+-}=0,\
L_{+i}=\frac{\mu}{4}\epsilon_{ijk}\delta F_{+jk-},\
L_{+b}=\frac{\mu}{2}\delta F_{123b}\nonumber\\& &L_{--}=0,\ \
L_{-i}=0,\ \ L_{-b}=0\nonumber\\&
&L_{ij}=\frac{\mu}{2}\delta_{ij}\delta F_{123-},\
L_{ib}=\frac{\mu}{4}\epsilon_{ijk}\delta F_{bjk-},\
L_{bd}=-\frac{\mu}{2}\delta _{bd}\delta F_{123-}\EE

Next let us deal with the Maxwell equation. In the absence of the
source, it is \BE
\label{eqn:maxwelleqn}\frac{1}{\sqrt{-g}}\partial_\lambda
\left(\sqrt{-g}\
F^{\lambda\mu_1\mu_2\mu_3}\right)-\frac{\tilde{\eta}}{1152}\frac{\epsilon^{\mu_1...\mu_{11}}}{\sqrt{-g}}
F_{\mu_4...\mu_7}F_{\mu_8...\mu_{11}}=0\EE where $\tilde{\eta}$ is
either $+1$ or $-1$ depending on the choice of convention, which one
can fix later by requiring the consistency of the conventions for
the equations and the solutions under consideration. When the source
is present, we add its current $J^{\mu_1\mu_2\mu_3}$ to the
left-hand side of the above equation, and get \BE
\delta\left[\frac{1}{\sqrt{-g}}\partial_\lambda \left(\sqrt{-g}\
F^{\lambda\mu_1\mu_2\mu_3}\right)-\frac{\tilde{\eta}}{1152}\frac{\epsilon^{\mu_1...\mu_{11}}}{\sqrt{-g}}
F_{\mu_4...\mu_7}F_{\mu_8...\mu_{11}}\right]=J^{\mu_1\mu_2\mu_3}\EE

We can write the left-hand side of the above equation as the sum of
two totally antisymmetric tensors
$Z^{\mu_1\mu_2\mu_3}+S^{\mu_1\mu_2\mu_3}$, where
$Z^{\mu_1\mu_2\mu_3}$ is defined to be the part that contains the
metric perturbation only, and $S^{\mu_1\mu_2\mu_3}$ is defined to be
the part that contains the three-form perturbation only. One finds
\BE& & Z^{+-i}=\mu\epsilon_{ijk}\partial_j\bar{h}_{-k},\ \
Z^{+-b}=0, \ \
Z^{+ij}=\mu\epsilon_{ijk}(\partial_-\bar{h}_{-k}-\partial_k\bar{h}_{--}),
\ \ Z^{+ib}=0,\ \ Z^{+bc}=0\nonumber\\&
&Z^{-ij}=\mu\epsilon_{ijk}\left[\partial_k\left(\frac{1}{3}\bar{h}
-\bar{h}_-^{\
-}-\sum_{i=1}^3\bar{h}_{ii}\right)-\partial_b\bar{h}_{kb}\right],\ \
Z^{-ib}=\mu\epsilon_{ijk}\partial_j\bar{h}_{kb},\ \
Z^{-bc}=0\nonumber\\&
&Z^{ijk}=-\mu\epsilon_{ijk}\left[\partial_-\left(\frac{1}{3}\bar{h}
-\bar{h}_-^{\
-}-\sum_{i=1}^3\bar{h}_{ii}\right)-\partial_b\bar{h}_{-b}\right], \
\
Z^{ijb}=\mu\epsilon_{ijk}(\partial_-\bar{h}_{kb}-\partial_k\bar{h}_{-b})\nonumber\\&
&Z^{ibc}=0,\ \ Z^{bce}=0\EE and \BE
S^{+-A}=g^{\mu\nu}\partial_\mu\partial_\nu a_{-+A} +\partial_B
g_{++}\partial_- a_{BA-}-\partial_-(\nabla^\mu
a_{\mu+A})+\partial_A(\nabla^\mu a_{\mu+-})\EE \BE
S^{+AB}=g^{\mu\nu}\partial_\mu\partial_\nu
a_{-AB}-\partial_-(\nabla^\mu a_{\mu AB})+\partial_A(\nabla^\mu
a_{\mu-B})-\partial_B(\nabla^\mu a_{\mu-A})\EE \BE S^{-AB}=&
&g^{\mu\nu}\partial_\mu\partial_\nu
a_{+AB}-g_{++}S^{+AB}\nonumber\\&
&+\left\{\left[(\partial_Ag_{++})(\partial_-a_{-+B})+\partial_A(\nabla^\mu
a_{\mu+B})-\partial_A\left(a_{EB-}\partial_Eg_{++}\right)\right]-\left[A\leftrightarrow
B\right]\right\}\nonumber\\& &-(\partial_Dg_{++})\delta
F_{D-AB}-\mu\frac{\tilde{\eta}}{24}\epsilon^{-AB\mu_4...\mu_7
123+}\delta F_{\mu_4...\mu_7}\EE \BE S^{ABE}=&
&g^{\mu\nu}\partial_\mu\partial_\nu
a_{ABE}-(\partial_Ag_{++})(\partial_-a_{-BE})-(\partial_Bg_{++})(\partial_-a_{-EA})
-(\partial_Eg_{++})(\partial_-a_{-AB})\nonumber\\&
&-\partial_A(\nabla^\mu a_{\mu BE})-\partial_B(\nabla^\mu a_{\mu
EA})-\partial_E(\nabla^\mu a_{\mu
AB})-\mu\frac{\tilde{\eta}}{24}\epsilon^{ABE\mu_4...\mu_7
123+}\delta F_{\mu_4...\mu_7}\nonumber\\
\EE Notice that $S^{\mu_1\mu_2\mu_3}$ contains $\nabla^\mu
a_{\mu\rho\lambda}$ and its derivatives. Those terms correspond to
the gauge freedom for the three-form gauge potential.

Now that we have collected the expressions for the various tensors,
we are ready to diagonalize the field equations. Recall that the
Einstein equation is \BE
-\frac{1}{2}\nabla^\sigma\nabla_\sigma\bar{h}_{\mu\nu}+K_{\mu\nu}+Q_{\mu\nu}
-N_{\mu\nu}-L_{\mu\nu}=\kappa_{11}^2[T_{\mu\nu}]_S\EE and the
Maxwell equation is \BE
Z^{\mu_1\mu_2\mu_3}+S^{\mu_1\mu_2\mu_3}=J^{\mu_1\mu_2\mu_3}\EE The
right-hand side of these equations is given by specifying the
source that we consider (recall that the three-form current $J$ is
of order $\kappa_{11}^2$), hence we only need to concentrate on
diagonalizing the left-hand sides.

As will be seen shortly, it is useful to define ``level'' for
tensors: lower $+$/upper $-$ indices contribute $+1$ to the level;
lower $-$/upper $+$ indices contribute $-1$ to level; and the upper
$A$/lower $A$ indices contribute zero to the level. We shall see
that the field equations should be solved in ascending order of
their levels. The following is the detailed prescription of the
diagonalization procedure. Let us use the shorthand notation
$(E.E.)_{\mu\nu}$ for the lower $(\mu\nu)$ component of the Einstein
equation, and $(M.E.)^{\mu_1\mu_2\mu_3}$ for the upper
$(\mu_1\mu_2\mu_3)$ component of the Maxwell equation.

$\bullet$\ \ at level $-2$

The only field equation at this level is $(E.E.)_{--}$, which reads,
upon using the expressions of the various tensors
$\nabla^\sigma\nabla_\sigma\bar{h}_{\mu\nu}, K_{\mu\nu},
Q_{\mu\nu}...$ etc., that we have given above
\begin{equation}\label{laplace_eqn_for_hmm}
-\frac{1}{2}g^{\mu\nu}\partial_\mu\partial_\nu\bar{h}_{--}+Q_{--}=\kappa_{11}^2[T_{--}]_S\end{equation}
This equation can be immediately solved for $\bar{h}_{--}$ after
specifying the source term and the gauge choice term $Q_{--}$.

$\bullet\ \ $ at level $-1$

We have $(E.E.)_{-A}$, which reads \BE
-\frac{1}{2}\left[g^{\mu\nu}\partial_\mu\partial_\nu\bar{h}_{-A}
-(\partial_Ag_{++})(\partial_-\bar{h}_{--})\right]+Q_{-A}=\kappa_{11}^2[T_{-A}]_S\EE
which can now be solved for $\bar{h}_{-A}$, using the $\bar{h}_{--}$
found previously. Also at this level is $(M.E.)^{+AB}$, which reads,
\BE& & g^{\mu\nu}\partial_\mu\partial_\nu
a_{-ij}-\partial_-(\nabla^\mu a_{\mu ij})+\partial_i(\nabla^\mu
a_{\mu -j})-\partial_j(\nabla^\mu a_{\mu
-i})+\mu\epsilon_{ijk}(\partial_-\bar{h}_{-k}-\partial_k\bar{h}_{--})=J^{+ij}\nonumber\\
& &g^{\mu\nu}\partial_\mu\partial_\nu a_{-ib}-\partial_-(\nabla^\mu
a_{\mu ib})+\partial_i(\nabla^\mu a_{\mu -b})-\partial_b(\nabla^\mu
a_{\mu -i})=J^{+ib}\nonumber\\& &g^{\mu\nu}\partial_\mu\partial_\nu
a_{-bc}-\partial_-(\nabla^\mu a_{\mu bc})+\partial_b(\nabla^\mu
a_{\mu -c})-\partial_c(\nabla^\mu a_{\mu -b})=J^{+bc}\EE from which
we can find $a_{-AB}$, upon specifying the gauge choice $\nabla^\mu
a_{\mu\rho\lambda}$ for the three-form and using the $\bar{h}_{-A}$
and $\bar{h}_{--}$ found previously.

$\bullet$\ \ at level $0$

At this level we have $(E.E.)_{+-}$, $(M.E.)^{+-A}$, $(E.E.)_{AB}$,
and $(M.E.)^{ABE}$.

$(E.E.)_{+-}$ is of the form \BE
-\frac{1}{2}g^{\mu\nu}\partial_\mu\partial_\nu\bar{h}_{+-}=\text{known
terms}\EE (From now on, we will not bother writing down the detailed
equations; ``known terms'' refers to the gauge choice terms
$Q_{\mu\nu}$, $\nabla^\mu a_{\mu\rho\lambda}$, source terms, and
terms containing previously found $\bar{h}_{\mu\nu}$'s and
$a_{\mu\nu\rho}$'s,  one can write those down by looking up the
expressions given earlier for the various tensors.) Hence solving it
we get $\bar{h}_{+-}$. Solving $(M.E.)^{+-A}$ gives $a_{-+A}$.

$(E.E.)_{AB}$ and $(M.E.)^{ABE}$ are coupled, so a little more work
is needed. The following are the details. First notice that the only
unknown in $(M.E.)^{bce}$ is $a_{bce}$, hence solving this equation
we find $a_{bce}$\ ($(M.E.)^{bce}$ contains the usual term
$g^{\mu\nu}\partial_\mu\partial_\nu a_{bce}$ and also a term of the
form $\partial_- a_{dfg}$ which comes from the $F\wedge F$ in the
Maxwell equation, hence it is not quite a Laplace equation. But,
that being said, one shouldn't have any difficulty solving it.)

$(M.E.)^{ibc}$ is of the form $g^{\mu\nu}\partial_\mu\partial_\nu
a_{ibc}=\text{known terms}$, solving which gives $a_{ibc}$.
$(M.E.)^{ijb}$ and $(E.E.)_{kb}$ are coupled in the following manner
\BE& & g^{\mu\nu}\partial_\mu\partial_\nu
a_{ijb}+\mu\epsilon_{ijk}\partial_-\bar{h}_{kb}=\text{known
terms}\nonumber\\&
&-\frac{1}{2}g^{\mu\nu}\partial_\mu\partial_\nu\bar{h}_{kb}
+\frac{1}{4}\mu\epsilon_{klm}\partial_-a_{lmb}=\text{known terms}\EE
Decoupling these two equations is quite easy. Let us take $a_{12b}$
and $\bar{h}_{3b}$ as the representative case. One sees that these
two equations can be recombined to give \BE& &
(g^{\mu\nu}\partial_\mu\partial_\nu+i\mu\partial_-)(\bar{h}_{3b}+ia_{12b})
=\text{known terms}\nonumber\\&
&(g^{\mu\nu}\partial_\mu\partial_\nu-i\mu\partial_-)(\bar{h}_{3b}-ia_{12b})
=\text{known terms} \EE Solving these equations gives
$(\bar{h}_{3b}+ia_{12b})$ and $(\bar{h}_{3b}-ia_{12b})$, and in turn
$\bar{h}_{3b}$ and $a_{12b}$.

$(M.E.)^{ijk}$ is coupled to $(E.E.)_{ij}$ and $(E.E.)_{bd}$ through
the quantity $H\equiv \frac{2}{3}\sum_{i=1}^3\bar{h}_{ii}
-\frac{1}{3}\sum_{a=4}^9\bar{h}_{aa}$ in the following manner \BE &
&g^{\mu\nu}\partial_\mu\partial_\nu
a_{123}+\mu\partial_-H=\text{known terms}\nonumber\\ &
&-\frac{1}{2}g^{\mu\nu}\partial_\mu\partial_\nu\bar{h}_{ij}
+\frac{1}{2}\mu\delta_{ij}\partial_-a_{123}=\text{known terms}\nonumber\\
& &-\frac{1}{2}g^{\mu\nu}\partial_\mu\partial_\nu\bar{h}_{bd}
-\frac{1}{2}\mu\delta_{bd}\partial_-a_{123}=\text{known terms}\EE
Combining the last two equations gives \BE
-g^{\mu\nu}\partial_\mu\partial_\nu
H+4\mu\partial_-a_{123}=\text{known terms} \EE Recombining this with
first equation, we get \BE& &
(g^{\mu\nu}\partial_\mu\partial_\nu+2i\mu\partial_-)(H+2ia_{123})=\text{known
terms}\nonumber\\ & &
(g^{\mu\nu}\partial_\mu\partial_\nu-2i\mu\partial_-)(H-2ia_{123})=\text{known
terms}\EE solving which individually gives $H$ and $a_{123}$. Using
the resulting expression for $a_{123}$ one can then find
$\bar{h}_{ij}$ and $\bar{h}_{bd}$. Thus we are done with
$(E.E.)_{AB}$ and $(M.E.)^{ABE}$.

$\bullet$\ \ at level $1$

$(M.E.)^{-AB}$ is of the form $g^{\mu\nu}\partial_\mu\partial_\nu
a_{+AB}=\text{known terms}$, solving which gives $a_{+AB}$.

$(E.E.)_{+A}$ is of the form
$-\frac{1}{2}g^{\mu\nu}\partial_\mu\partial_\nu\bar{h}_{+A}
=\text{known terms}$, solving which gives $\bar{h}_{+A}$.

$\bullet$\ \ at level $2$

$(E.E.)_{++}$ is of the form
$-\frac{1}{2}g^{\mu\nu}\partial_\mu\partial_\nu\bar{h}_{++}
=\text{known terms}$, solving which gives $\bar{h}_{++}$.
\newline
\newline
Thus we have diagonalized the whole set of Einstein equations and
Maxwell equations.

\subsection{Application to a Spherical Membrane Source using the Near-Membrane Expansion}
Now let us apply the general formalism of the previous subsection to
the case of interest, with the source being a spherical membrane
sitting at the origin of the transverse directions, i.e. having
$(X^1)^2+(X^2)^2+(X^3)^2=r_0^2$, $X^4=0,...,X^9=0$, and
$X^+=t,X^-=0$. The gauge choice we shall take is: $q_\alpha=0$
(hence all the $Q_{\alpha\beta}$'s vanish); and $\nabla^\mu
a_{\mu\rho\lambda}=0$. The nonzero components of the stress tensor
and three-form current for this source are given by \BE &
&[T_{--}]_s=T\delta(x^-)\delta(r-r_0)\delta(x^4)...\delta(x^9)
\left(\frac{\mu r_0}{3}\right)^{-1}\nonumber\\
& &[T_{+-}]_s=-\left(\frac{\mu r_0}{3}\right)^2[T_{--}]_s,\ \
[T_{ij}]_s=\left(\frac{\mu
r_0}{3}\right)^2\left(\frac{x^ix^j}{r_0^2}-\delta^{ij}\right)[T_{--}]_s,\
\ [T_{++}]_s=\left(\frac{\mu
r_0}{3}\right)^4[T_{--}]_s\nonumber\\
\EE and \BE J^{+ij}=\kappa_{11}^2(-2)\left(\frac{\mu
r_0}{3}\right)\epsilon_{ijk}\frac{x^k}{r_0}[T_{--}]_s\EE where
$r\equiv\sqrt{x^ix^i}$.

Now let us explain what we mean by ``near-membrane expansion''.
Define $w\equiv r-r_0$, $z\equiv \sqrt{x^ax^a}$, and
$\xi\equiv\sqrt{w^2+z^2}$, which parameterize the distance away from
the source membrane. We shall assume that $w,z,\xi$ are of the same
order of magnitude. The near-membrane expansion is an expansion in
$\xi/r_0$. When one sits very close to membrane, one just sees a
flat membrane, which is the zeroth order of the expansion. As one
moves away from the membrane, one begins to feel the curvature of
membrane, which gives the higher order corrections in the expansion.
One should also note that the zeroth order of this expansion is just
the flat space limit: $\mu\rightarrow 0$, $r_0\rightarrow \infty$,
with $\mu r_0$ kept finite.

It is instructive to see how the zeroth order works. At this order,
in the Einstein equations and Maxwell equations, the effective
source terms (which are of the forms $(\partial
g_{++})\bar{h}_{\mu\nu}$, etc.) arising from the various tensors
$K_{\mu\nu}$, $N_{\mu\nu}$, $L_{\mu\nu}$, $Z^{\mu\nu\rho}$ etc. are
less singular then the delta-functional sources $[T_{\mu\nu}]_s$ and
$J^{\mu\nu\rho}$, and can thus be thrown away. Then the resulting
equations are trivially decoupled. Also, at this order, we can treat
the $x^i$'s in $[T_{\mu\nu}]_s$ and $J^{\mu\nu\rho}$ as constant
vectors. One finds (using the subscript $0$ to denote ``zeroth
order'') \BE & &[\bar{h}_{-A}]_0=0,\ \ [a_{-ij}]_0=\left(\frac{\mu
r_0}{3}\right)
\epsilon_{ijk}\frac{x^k}{r_0}[\bar{h}_{--}]_0,\ \  [a_{-ib}]_0=0,\ \ [a_{-bc}]_0=0\nonumber\\
& &[\bar{h}_{+-}]_0=-\left(\frac{\mu
r_0}{3}\right)^2[\bar{h}_{--}]_0,\ \ [a_{-+A}]_0=0, \ \
[\bar{h}_{ij}]_0=\left(\frac{\mu
r_0}{3}\right)^2\left(\frac{x^ix^j}{r_0^2}-\delta^{ij}\right)[\bar{h}_{--}]_0\nonumber\\
& &[\bar{h}_{ib}]_0=0,\ \ [\bar{h}_{bc}]_0=0,\ \
[a_{ABE}]_0=0\nonumber\\
& &[a_{+ij}]_0=-\left(\frac{\mu
r_0}{3}\right)^3\epsilon_{ijk}\frac{x^k}{r_0}[\bar{h}_{--}]_0,\ \
[a_{+ib}]_0=0,\ \ [a_{+bc}]_0=0,\ \ [\bar{h}_{+A}]_0=0\nonumber\\
& &[\bar{h}_{++}]_0=\left(\frac{\mu
r_0}{3}\right)^4[\bar{h}_{--}]_0\EE where $[\bar{h}_{--}]_0$
satisfies \BE \left[ -\left(\frac{\mu
r_0}{3}\right)^2k_-^2+\frac{\partial^2}{(\partial
w)^2}+\frac{\partial^2}{(\partial
x^4)^2}+...\frac{\partial^2}{(\partial
x^9)^2}\right][\bar{h}_{--}]_0=-\frac{1}{\pi R}\left(\frac{\mu
r_0}{3}\right)^{-1}\kappa_{11}^2
T\delta(w)\delta(x^4)...\delta(x^9)\nonumber\\
\EE (where we have
multiplied the right hand side of the equation by $\frac{1}{2\pi R}$
due to the Fourier transform along the $x^-$ direction), and is
given by \BE [\bar{h}_{--}]_0=\Delta\frac{\exp\left(-\ \frac{\mu
r_0}{3}k_-\xi\right)}{\xi^5} \left[3+3\left(\frac{\mu
r_0}{3}k_-\xi\right)+\left(\frac{\mu
r_0}{3}k_-\xi\right)^2\right]\EE with $\Delta\equiv
\frac{\kappa_{11}^2T}{16\pi^4R\left(\frac{\mu r_0}{3}\right)}$.

Plugging the above zeroth order solution of the field equations into
the light-cone Lagrangian $\delta {\cal L}_{lc}$ given in eqn
(\ref{eqn:deltaPi+}), for a spherical probe membrane with radius
$r_0'$, sitting at rest in the $1,2,3$ directions, and moving about
in the $4$ through $9$ directions: $(X^1)^2+(X^2)^2+(X^3)^2=r_0'^2$,
$X^4(t),...,X^9(t)$, and $X^+=t,X^-=0$, one finds (using the facts
that $T=\frac{\mu\Pi_-}{3r_0'\sin\theta}$, and we can set $r_0'=r_0$
because we are looking at the zeroth order) \BE
\label{eqnLlc0}\delta {\cal
L}_{lc}=\frac{1}{8}\Pi_-[\bar{h}_{--}]_0(\dot{X}^a\dot{X}^a)^2\EE

It is worth noting that, keeping only the leading order term in
$k_-$ in $[\bar{h}_{--}]_0$, eqn (\ref{eqnLlc0}) becomes the $v^4$
Lagrangian for the case of longitudinal momentum transfer between
two membranes in the flat space, given in \cite{Polchinski:1997pz}.

Now let us go on to consider higher orders in the near-membrane
expansion. Since in this paper we do not consider longitudinal
momentum transfer, we shall set $k_-=0$ (which makes many fields
equations decouple).

Denote \BE
\Box&\equiv&\partial_A\partial_A\nonumber\\
& &\text{(when acting on functions of $(w,z))$}\nonumber\\
&=&\Box_0+\delta\Box \EE with
$\Box_0\equiv\frac{\partial^2}{\partial
w^2}+\frac{\partial^2}{\partial
z^2}+\frac{5}{z}\frac{\partial}{\partial z}
=\partial_w^2+\partial_a\partial_a$ being the zeroth order Laplace
operator, and $\delta\Box\equiv
\frac{2}{r_0+w}\frac{\partial}{\partial w}$ being curvature
correction to it.

{\bf At level $-2$}
\\
$(E.E.)_{--}$ \BE \Box\bar{h}_{--}=-\frac{1}{\pi
R}\kappa_{11}^2T\delta(w)\delta(x^4)...\delta(x^9)\left(\frac{\mu
r_0}{3}\right)^{-1} \EE Let \BE
\bar{h}_{--}=[\bar{h}_{--}]_0+\delta\bar{h}_{--} \EE with
$[\bar{h}_{--}]_0\equiv \Delta\frac{3}{\xi^5}$, which satisfies \BE
\Box_0[\bar{h}_{--}]_0=-\frac{1}{\pi
R}\kappa_{11}^2T\delta(w)\delta(x^4)...\delta(x^9)\left(\frac{\mu
r_0}{3}\right)^{-1}\EE Then $(E.E.)_{--}$ becomes \BE
\Box_0\delta\bar{h}_{--}+\delta\Box
[\bar{h}_{--}]_0+\delta\Box\delta\bar{h}_{--}=0 \EE Now let's look
at the order of magnitude of each term in the above equation. Notice
that $\Box_0\sim \frac{1}{\xi^2}$, $\delta\Box\sim
\frac{1}{r_0\xi}$. The second term is thus $\sim
\Delta\frac{1}{r_0\xi^6}$, which tells us
$\delta\bar{h}_{--}\sim\Delta\frac{1}{r_0\xi^4}$. Solving the
equation iteratively we find \BE
\delta\bar{h}_{--}=[\bar{h}_{--}]_0\left(-\frac{w}{r_0}
+\frac{w^2}{r_0^2}-\frac{w^3}{r_0^3}+\frac{w^4}{r_0^4}\right) \EE
and thus \BE
\bar{h}_{--}=3\Delta\frac{1}{\xi^5}\left[\frac{r_0}{r_0+w}+O\left(\frac{\xi^5}{r_0^5}\right)\right]
\EE We did not compute the $O\left(\frac{\xi^5}{r_0^5}\right)$
terms, because we are only interested in the part of the solution
that is singular as $\xi\rightarrow 0$. Solving the other field
equations is similar, so we just present the results below, omitting
the $O\left(\frac{\xi^5}{r_0^5}\right)$ symbol.
\\
At level $-1$ \BE
& &\bar{h}_{-A}=0\nonumber\\
& &a_{-bc}=0\nonumber\\
& &a_{-ib}=0\nonumber\\
&
&a_{-ij}=\epsilon_{ijk}x^k\mu\Delta\frac{1}{\xi^5}\left[1-\frac{1}{2}\frac{w}{r_0}+\frac{1}{6}\frac{w^2+z^2}{r_0^2}
-\frac{1}{2}\frac{wz^2}{r_0^3}+\frac{w^2z^2}{r_0^4}\right] \EE At
level $0$ \BE & &\bar{h}_{+-}=-\left(\frac{\mu
r_0}{3}\right)^2\frac{3\Delta}{\xi^5}\left[1+\left(\frac{5}{4}\frac{w^2}{r_0^2}
+\frac{7}{8}\frac{z^2}{r_0^2}\right)\frac{r_0}{r_0+w}\right]\nonumber\\
& &a_{-+A}=0\nonumber\\
& &a_{ABD}=0\nonumber\\
\EE \BE
\bar{h}_{kb}=x^kx^b\mu^2\Delta\frac{1}{\xi^5}\left(-\frac{1}{2}\right)\left[1-\frac{5}{4}\frac{w}{r_0}
+\frac{17}{12}\frac{w^2}{r_0^2}-\frac{1}{12}\frac{z^2}{r_0^2}-\frac{3}{2}\frac{w^3}{r_0^3}+\frac{1}{4}\frac{wz^2}{r_0^3}
+\frac{3}{2}\frac{w^4}{r_0^4}-\frac{1}{2}\frac{w^2z^2}{r_0^4}\right]
\EE \BE \bar{h}_{ij}=& &\frac{x^ix^j}{r^2}\left(\frac{\mu
r_0}{3}\right)^2\Delta\frac{3}{\xi^5}\left[1-\frac{w}{r_0}-\frac{z^2}{r_0^2}
+\frac{w^3}{r_0^3}+2\frac{wz^2}{r_0^3}-\frac{w^4}{r_0^4}-\frac{w^2z^2}{r_0^4}+\frac{z^4}{r_0^4}\right]\nonumber\\
& &-\delta^{ij}\left(\frac{\mu
r_0}{3}\right)^2\Delta\frac{3}{\xi^5}\left[1+\frac{1}{2}\frac{w}{r_0}+\frac{7}{12}\frac{w^2}{r_0^2}
-\frac{1}{24}\frac{z^2}{r_0^2}-\frac{1}{4}\frac{w^3}{r_0^3}+\frac{3}{8}\frac{wz^2}{r_0^3}+\frac{1}{4}\frac{w^4}{r_0^4}
-\frac{1}{24}\frac{w^2z^2}{r_0^4}+\frac{1}{3}\frac{z^4}{r_0^4}\right]\nonumber\\
\EE \BE \bar{h}_{bd}=\delta_{bd}(\mu
r_0)^2\Delta\frac{1}{\xi^5}\left[\frac{1}{2}\frac{w}{r_0}-\frac{7}{18}\frac{w^2}{r_0^2}
-\frac{19}{72}\frac{z^2}{r_0^2}+\frac{7}{18}\frac{w^3}{r_0^3}+\frac{19}{72}\frac{wz^2}{r_0^3}-\frac{7}{18}\frac{w^4}{r_0^4}
-\frac{19}{72}\frac{w^2z^2}{r_0^4}\right] \EE At level $+1$ \BE
& &a_{+bc}=0\nonumber\\
& &a_{+ib}=0\nonumber\\
& &a_{+ij}=-\epsilon_{ijk}x^k\left(\frac{\mu
r_0}{3}\right)^2\mu\Delta\frac{1}{\xi^5}\left[1+2\frac{w}{r_0}-\frac{w^2}{r_0^2}
-\frac{5}{4}\frac{z^2}{r_0^2}+\frac{3}{2}\frac{w^3}{r_0^3}+\frac{7}{4}\frac{wz^2}{r_0^3}-\frac{3}{2}\frac{w^4}{r_0^4}
-\frac{5}{4}\frac{w^2z^2}{r_0^4}+\frac{1}{2}\frac{z^4}{r_0^4}\right]\nonumber\\
\EE \BE \bar{h}_{+A}=0 \EE At level $+2$ \BE \bar{h}_{++}=
\left(\frac{\mu
r_0}{3}\right)^4\Delta\frac{3}{\xi^5}\left[1+\frac{5}{2}\frac{w}{r_0}+\frac{31}{12}\frac{w^2}{r_0^2}
-\frac{1}{24}\frac{z^2}{r_0^2}+\frac{17}{12}\frac{w^3}{r_0^3}-\frac{1}{12}\frac{wz^2}{r_0^3}+\frac{1}{3}\frac{w^4}{r_0^4}
+\frac{23}{24}\frac{w^2z^2}{r_0^4}+\frac{17}{32}\frac{z^4}{r_0^4}\right]\nonumber\\
\EE (A note in passing: as it turns out, the choice of convention
for $\tilde{\eta}$ in the Maxwell equation (\ref{eqn:maxwelleqn})
does not matter, since it drops out when solving the field equations
for our specific source.)

Again, let the probe membrane have a radius $r_0'=r_0+w$, with the
trajectory $(X^1)^2+(X^2)^2+(X^3)^2=r_0'^2$, $X^4(t),...,X^9(t)$,
and $X^+=t,X^-=0$. We shall take $v\equiv\sqrt{\dot{X}^a\dot{X}^a}$
to be of order $\mu z$ (recall that the supersymmetric circular
orbit has $v=\frac{1}{6}\mu z$; so here we are considering
generically nonsupersymmetric orbits that can be regarded as
deformations of the supersymmetric circular one). Plugging in the
supergravity solution given above into eqn (\ref{eqn:deltaPi+} ),
using $T=\frac{\mu\Pi_-}{3r_0'\sin\theta}$, and in the end keeping
only the part of $\delta{\cal L}_{lc}$ that is singular as
$\xi\rightarrow 0$, we find the probe's ${\cal L}_{lc}$ to be \BE
{\cal L}_{lc}=({\cal L}_{lc})_{pp}+\delta {\cal L}_{lc}\EE with
$({\cal L}_{lc})_{pp}$ being the action in the unperturbed pp-wave
background, and \BE \delta {\cal L}_{lc}=
\Pi_-\Delta\frac{\mu^4(4w^2z^2+7z^4)-72\mu^2v^2(2w^2+5z^2)+3888v^4}{10368\xi^5}\EE
 Notice that the above $\delta{\cal L}_{lc}$'s singular behavior as $\xi\rightarrow 0$
 is homogeneous: $\sim \frac{1}{\xi}$ (since $v$ is of order $\mu z$).  The expression in
the numerator: $\mu^4(4w^2z^2+7z^4)-72\mu^2v^2(2w^2+5z^2)+3888v^4$
nicely factorizes into $(36v^2-z^2\mu^2)[108v^2-(4w^2+7z^2)\mu^2]$,
which shows that for the special case of the supersymmetric circular
orbit $v=\frac{1}{6}\mu z$ considered in \cite{Shin:2003np},
$\delta{L}_{lc}$ vanishes as expected.

To be more precise, so far we have been talking about Lagrangian
density. Since the membrane worldvolume is taken to be a unit
sphere, the Lagragian is given by \BE\label{eqn:deltaLlc_result}
\delta L_{lc}=\int d\theta d\phi \delta{\cal
L}_{lc}&=&\frac{\kappa_{11}^2T^2}{4\pi^3R\mu^2}\frac{(36v^2-z^2\mu^2)[108v^2-(4w^2+7z^2)\mu^2]}{1152\xi^5}\nonumber\\
&=&\frac{\alpha}{\mu^2}\frac{(36v^2-z^2\mu^2)[108v^2-(4w^2+7z^2)\mu^2]}{1152\xi^5}
\EE where to get the first line we used
$T=\frac{\mu\Pi_-}{3r_0'\sin\theta}$ to eliminate $\Pi_-$ in terms
of $T$ and $r_0'$, and set $r_0' \approx r_0$ in the end to remove higher $\frac{w}{r_0}$ order curvature correction. To
get the second line we used the expressions for $\kappa_{11}^2$,
$T$, and $\alpha$ in terms of $M$ and $R$ given at the end of
Subsection \ref{subsection:sphericalmembrane}.

Here we would like to make a brief comparison of the above membrane
result with the graviton result given in \cite{Lee:2003kf}.

First of all, the membrane result contains the variable $w$ (the
difference in radius between the probe membrane and the source membrane), which has no counterpart in the graviton case. Secondly, in
terms of the $x^1$ through $x^3$ directions, the two membranes are
sitting at rest at the origin; this corresponds to setting $x^i=0$
and $v_i=0$ in the graviton case.

If we set $w=0$ in eqn (\ref{eqn:deltaLlc_result}), i.e., consider
two membranes of the same size, then $\xi=z$, and
\BE\label{eqn:deltaLcmembrane} (\delta L_{lc})_{\text{membrane}} =
\frac{\alpha}{1152\mu^2z^5}(36v^2-z^2\mu^2)(108v^2-7z^2\mu^2)\EE
while for the gravitons, upon setting $N_p=N_s=N$, $x^i=0$, and
$v_i=0$ in eqn (19) of \cite{Lee:2003kf}, we have
\BE\label{eqn:deltalcgraviton} (\delta
 L_{lc})_{\text{graviton}}=\frac{\alpha^3N^2}{5736z^7}(36v^2-z^2\mu^2)(140v^2-7z^2\mu^2)\EE
When comparing the above two expressions, note the difference
between the numerators: $(108v^2-7z^2\mu^2)$ for membrane, and
$(140v^2-7z^2\mu^2)$ for graviton. Also notice their different power
law dependence on $z$: $\frac{1}{z^5}$ for membrane and
$\frac{1}{z^7}$ for graviton, which cannot be undone by integrating
over the membrane ($r_0$, the radius of the membrane, has
nothing to do with $z$, the separation of the membranes in the $X^4$ to $X^9$ directions).


\section{Matrix Theory Computation - The Membrane Limit}\label{gaugeside}
Shin and Yoshida have previously calculated the one-loop effective
action for membrane fuzzy spheres extended in the first three
directions and having periodic motion in a sub-plane of the
remaining six transverse directions. In reference
 \cite{Shin:2003np} they considered the case of supersymmetric circular motion for an 
arbitrary radius and angular frequency $\frac{\mu}{6}$ (this
orbit preserves eight supersymmetries and was first found by
\cite{Berenstein:2002jq}). Here we generalize that analysis to
 orbits which are not required to satisfy the classical
equations of motion and are in general nonsupersymmetric. The
procedures are: expanding the action to quadratic order in
fluctuations, writing the fields in terms of the matrix spherical
harmonics introduced in \cite{Dasgupta:2002hx} \footnote{For the
computation below, one only needs the transformation of the matrix
spherical harmonics under $SU(2)$, however, for detailed
construction of the matrix spherical harmonics, see, e.g., Appendix
A of \cite{Das:2003yq}.} , diagonalizing the mass matrices of the
bosons, fermions, and ghosts, and finally summing up the masses to
get the one-loop effective action. In doing so, we shall adopt the
notations of \cite{Shin:2003np}. We shall consider the background
\be B^{I}=\left(
\begin{array}{cc} B^I_{(1)}&0\\0&B^I_{(2)}
\end{array} \right) \ee where \be
B^i_{(1)}=\frac{\mu}{3}J^i_{(1)N_1 \times N_1} \hspace{1cm} B^a_{(1)}= 0\cdot \  1_{N_1\times N_1}\nonumber\\
B^i_{(2)}=\frac{\mu}{3}J^i_{(2)N_2 \times N_2} + x^i(t) 1_{N_2\times
N_2}\hspace{1cm} B^a_{(2)}= x^a(t) 1_{N_2\times N_2} \ee with the
$J^i$'s being $su(2)$ generators. The above background has the
interpretation of one spherical membrane (labeled by the subscript
$(1)$) sitting at the origin, and the other spherical membrane
(labeled by the subscript $(2)$) moving along the arbitrary orbit
given by $\{x^i(t),x^a(t)\}$.

The fluctuations of the bosonic fields are given by \be
A=\left( \begin{array}{cc} Z^0_{(1)}& \Phi^0\\(\Phi^0)^{\dagger}&Z^0_{(2)} \end{array} \right)\nonumber\\
Y^{I}=\left( \begin{array}{cc} Z^I_{(1)}&\Phi^I\\
(\Phi^I)^{\dagger}&Z^I_{(2)} \end{array} \right) \ee As it turns
out, the part of the bosonic action containing the diagonal
fluctuations
 $Z^0,Z^I$ does not contain any new terms in addition to those given in
\cite{Shin:2003np}. So we do not write it out here. (It shall be the
same situation for the fermionic and ghost parts of the action; it
 is the off-diagonal fluctuations that give the one-loop interaction potential between the two membranes.)
 The action for the off-diagonal fluctuations is \be S_{OD}=\int dt\hspace{1mm}Tr\hspace{-3mm}
 &\left[\right.& \hspace{-3mm}- \left.\dot{\Phi^0}|^2+x^2|\Phi^0|^2
        +\left(\frac{\mu}{3} \right)^2|J^i\circ\Phi^0|^2
        -2\left(\frac{\mu}{3}\right)x^iRe((J^i\circ\Phi^0)(\Phi^0)^{\dagger})\right.\nonumber\\
         &&\hspace{-3mm}+|\dot{\Phi^i}|^2-x^2|\Phi^i|^2
        -\left(\frac{\mu}{3}\right)^2|\Phi^i+i\epsilon^{ijk}J^j\circ\Phi^k|^2
        +\left(\frac{\mu}{3}\right)^2|J^i\circ\Phi^i|^2\nonumber\\
        &&\hspace{-3mm}+2\left(\frac{\mu}{3}\right)x^iRe((J^i\circ\Phi^0)(\Phi^0)^{\dagger})
        -2i\dot{x^i}((\Phi^0)^{\dagger}\Phi^i-(\Phi^i)^{\dagger}\Phi^0)\nonumber\\
        &&\hspace{-3mm}+ |\dot{\Phi^a}|^2-\left(x^2+\frac{\mu^2}{6^2}\right)|\Phi^a|^2
        -\left(\frac{\mu}{3}\right)^2|J^i\circ\Phi^a|^2\nonumber\\
        &&\hspace{-3mm}+\left. 2\left(\frac{\mu}{3}\right)x^iRe((J^i\circ\Phi^a)(\Phi^a)^{\dagger})
        -2i\dot{x^a}((\Phi^0)^{\dagger}\Phi^a-(\Phi^a)^{\dagger}\Phi^0)\right]
\ee where $x^2\equiv x^ix^i+x^ax^a$, and dots means
time-derivatives.

We specialize to the case where, $x^i=0,x^8=b,x^9=vt$, with
$(x^8)^2+(x^9)^2$ denoted by $z^2$. The effective potential will be
computed by summing over the mass of the fermionic and ghost
fluctuations and then subtracting the mass of the bosonic
fluctuations. This method, which we will refer to as the sum over mass method, is
the same as the one used in \cite{Lee:2003kf}. Although the above
trajectory has the form of a straight line with constant velocity in
the $(x^8,x^9)$ plane, the final expression of $V_{eff}$ in terms of
$z\equiv\sqrt{(x^a)^2}$ and $v\equiv\sqrt{(\dot{x^a})^2}$ should
suffice for the purpose of comparing with supergravity for arbitrary
orbits $x^a(t)$. One may ask whether the sum over mass formula is
valid when the masses of the fluctuations are time-dependent (one
origin for such a time-dependence is the acceleration of the
trajectory). In Section 4.2 and Appendix A of \cite{Lee:2003kf}, it
was carefully shown that, in the case of two-graviton interaction in
the pp-wave background, the sum over mass formula was sufficient in
computing the terms that could occur on the supergravity side. The
time-dependence in the masses of the fluctuations will give terms of
the form of matrix theory corrections to supergravity (i.e., terms
that dominate at extreme short distances and cannot be observed in
supergravity), which does not concern us since we are only interested
in a comparison with supergravity.  Here we expect a similar
argument to hold in the case of two-membrane interaction. The
rather non-trivial agreement with supergravity presented at the end
of this section and also the agreement with the work of Shin and Yoshida
\cite{Shin:2004az} in section \ref{subsection:comparison_with_shin} confirm the validity of
 the sum over mass method.

 Expand the fields in terms of matrix
spherical harmonics \be
\Phi^{0,I}=\sum^{\frac{1}{2}(N_1+N_2)-1}_{j=\frac{1}{2}|N_1-N_2|}
\sum^{j}_{m=-j}\phi^{0,I}_{jm} Y^{N_1\times N_2}_{jm}. \ee
 For our choice of background the masses of modes in the $i=1,2,3$ and $a=4,5,6,7,8$ directions
are the same as those in \cite{Shin:2003np}. For the gauge field and
$a=9$ direction we have
\be
 S=\int dt \sum^{\frac{1}{2}(N_1+N_2)-1}_{j=\frac{1}{2}|N_1-N_2|}\hspace{-5mm}&&\Bigl[-|\dot{\phi^0_{jm}}|^2 +\left(z^2+\left(\frac{\mu}{3}\right)^2j(j+1)\right)|\phi^0_{jm}|^2 \nonumber\\
&&\hspace{1mm}+|\dot{\phi^9_{jm}}|^2-\left(z^2+\left(\frac{\mu}{6}\right)^2+\left(\frac{\mu}{3}\right)^2j(j+1)\right)|\phi^0_{jm}|^2\nonumber\\
&&\hspace{1mm}-2iv\left(\left(\phi_{jm}^0\right)^{*}\phi_{jm}^9-\left(\phi_{jm}^9\right)^{*}\phi_{jm}^0\right)\Bigr]
\ee where there is an implicit sum over $-j\leq m\leq j$. It is
straightforward to diagonalize the mass matrix for these modes.
Combining all contributions from bosonic fluctuations we get an
bosonic effective potential\footnote{Note that our convention
differs from that of \cite{Shin:2003np} by an overall minus sign.
See section \ref{MTSugraIntro}.} given by, \be\label{eqn:VBeff}
V^B_{eff}=&&\hspace{-3mm}- \sum^{\frac{1}{2}(N_1+N_2)-2}_{j=\frac{1}{2}|N_1-N_2|-1}(2j+1) \sqrt{z^2+\left(\frac{\mu}{3}\right)^2 (j+1)^2}\nonumber \\
       &&\hspace{-3mm}-\sum^{\frac{1}{2}(N_1+N_2)}_{j=\frac{1}{2}|N_1-N_2|+1}(2j+1) \sqrt{z^2+\left(\frac{\mu}{3}\right)^2 j^2}\nonumber\\
       & &\hspace{-3mm}-\sum^{\frac{1}{2}(N_1+N_2)-1}_{j=\frac{1}{2}|N_1-N_2|}(2j+1) \sqrt{z^2+\left(\frac{\mu}{3}\right)^2 j(j+1)}\nonumber \\
       &&\hspace{-3mm}-\sum^{\frac{1}{2}(N_1+N_2)-1}_{j=\frac{1}{2}|N_1-N_2|}5(2j+1) \sqrt{z^2+\left(\frac{\mu}{3}\right)^2 (j+\frac{1}{2})^2}\nonumber \\
       & &\hspace{-3mm}-\sum^{\frac{1}{2}(N_1+N_2)-1}_{j=\frac{1}{2}|N_1-N_2|}(2j+1) \left( \sqrt{z^2+\left(\frac{\mu}{3}\right)^2 (j(j+1)+\frac{1}{8})+ \frac{1}{2}\sqrt{\left(\frac{\mu}{6}\right)^4+16v^2}}\right.\nonumber\\
       & &\hspace{3cm}\left. +\sqrt{z^2+\left(\frac{\mu}{3}\right)^2 (j(j+1)+\frac{1}{8})
             -\frac{1}{2}\sqrt{ \left(\frac{\mu}{6}\right)^4+16v^2}}\right)
\ee

Now for the fermions we start with the action given in
\cite{Shin:2003np}

\be
L_F=Tr(i\Psi^{\dagger}\dot{\Psi}-\Psi^{\dagger}\gamma^I[\Psi,B^I]-i\frac{\mu}{4}\Psi^{\dagger}\gamma^{123}\Psi)
\ee with \be \Psi=\left( \begin{array}{cc} \Psi_{(1)}&
\chi\\\chi^{\dagger}&\Psi_{(2)} \end{array} \right). \ee Again the
action for the diagonal fluctuations has no new terms, and for the
off-diagonal  part we decompose the $SO(9)$ spinor $\chi$
using the subgroup $SO(3)\times SO(6)\sim SU(2)\times SU(4)$
preserved by PP-wave, $\textbf{16}\rightarrow
(\textbf{2},\textbf{4})+(\bar{\textbf{2}},\bar{\textbf{4}})$ \be
\chi=\frac{1}{\sqrt{2}}\left(
\begin{array}{c} \chi_{A\alpha}\\\hat{\chi}^{A_{\alpha}}\end{array}
\right). \ee Substituting this into $L_F$, \be
L_F=Tr\hspace{-3mm}&\Bigl[&\hspace{-3mm}i(\chi^{\dagger})^{A\alpha}\dot{\chi}_{A\alpha}-\frac{1}{4}(\chi^{\dagger})^{A\alpha}\chi_{A\alpha}
-\frac{\mu}{3}(\chi^{\dagger})^{A\alpha}(\sigma^i)_{\alpha}^{\beta}J^i\circ\chi_{A\beta}\nonumber\\
&&\hspace{-3mm}+i(\hat{\chi}^{\dagger})^{A\alpha}\dot{\hat{\chi}}_{A\alpha}+\frac{1}{4}(\hat{\chi}^{\dagger})^{A\alpha}\hat{\chi}_{A\alpha}
+\frac{\mu}{3}(\hat{\chi}^{\dagger})^{A\alpha}(\sigma^i)_{\alpha}^{\beta}J^i\circ\hat{\chi}_{A\beta}\nonumber\\
&&\hspace{-3mm}+x^i(-(\chi^{\dagger})^{A\alpha}(\sigma^i)_{\alpha}^{\beta}\chi_{A\beta}+(\hat{\chi}^{\dagger})^{A\alpha}(\sigma^i)_{\alpha}^{\beta}\hat{\chi}_{A\beta})
 + x^a( (\chi^{\dagger})^{A\alpha} \rho^a_{AB}
\hat{\chi}^B_{\alpha} + (\hat{\chi}^{\dagger})_A^{\alpha}
((\rho^a)^{AB})^{\dagger} \chi_{B\alpha}) \Bigr]\nonumber \\
\ee with the $\sigma^i$'s and $\rho^a$'s being the gamma matrices
for $SO(3)$ and $SO(6)$ respectively. We now substitute our specific
background and expand in modes, \be
L_F&=&\sum^{\frac{1}{2}(N_1+N_2)-3/2}_{j=\frac{1}{2}|N_1-N_2|-1/2}\left[
i\pi_{jm}^{\dagger}\dot{\pi}_{jm}+i\hat{\pi}_{jm}^{\dagger}\dot{\hat{\pi}}_{jm}
   -\frac{1}{3}\left(j+\frac{3}{4}\right)(\pi_{jm}^{\dagger}\pi_{jm}-\hat{\pi}_{jm}^{\dagger}\hat{\pi}_{jm})\right.\nonumber\\
   &&\left.\hspace{3cm}+x^a(\pi^{\dagger}_{jm}\rho^a\hat{\pi}_{jm}+\hat{\pi}^{\dagger}_{jm}(\rho^a)^{\dagger}\pi_{jm})\right]\nonumber\\
   &&+\sum^{\frac{1}{2}(N_1+N_2)-1/2}_{j=\frac{1}{2}|N_1-N_2|+1/2}\left[ i\eta_{jm}^{\dagger}\dot{\eta}_{jm}+i\hat{\eta}_{jm}^{\dagger}\dot{\hat{\eta}}_{jm}
   -\frac{1}{3}\left(j+\frac{1}{4}\right)(\eta_{jm}^{\dagger}\eta_{jm}-\hat{\eta}_{jm}^{\dagger}\hat{\eta}_{jm})\right.\nonumber\\
   &&\left.\hspace{3cm}+x^a(\eta^{\dagger}_{jm}\rho^a\hat{\eta}_{jm}+\hat{\eta}^{\dagger}_{jm}(\rho^a)^{\dagger}\eta_{jm})\right]
\ee where again there is an implicit sum over $m$. This can be
diagonalized and contributes to the effective action \be
V^F_{eff}=&&2\sum^{\frac{1}{2}(N_1+N_2)-3/2}_{j=\frac{1}{2}|N_1-N_2|-1/2}(2j+1) (\sqrt{z^2+\left(\frac{\mu}{3}\right)^2 (j+\frac{3}{4})^2+v}+\sqrt{z^2+\left(\frac{\mu}{3}\right)^2 (j+\frac{3}{4})^2-v}\nonumber\\
       &&+2\sum^{\frac{1}{2}(N_1+N_2)-1/2}_{j=\frac{1}{2}|N_1-N_2|+1/2}(2j+1) (\sqrt{z^2+\left(\frac{\mu}{3}\right)^2 (j+\frac{1}{4})^2+v}
         +\sqrt{z^2+\left(\frac{\mu}{3}\right)^2
         (j+\frac{1}{4})^2-v}\nonumber\\
\ee There is also the contribution from the ghosts, however, this
has no new terms for our choice of background, \be
V^G_{eff}=2\sum^{\frac{1}{2}(N_1+N_2)-1}_{j=\frac{1}{2}|N_1-N_2|}(2j+1)
\sqrt{z^2+\left(\frac{\mu}{3}\right)^2 j(j+1)} \ee and the total
effective action is the sum of the three pieces \be
V_{eff}=V^B_{eff}+V^F_{eff}+V^G_{eff}.\ee
\noindent

We introduce the variables $N$ and $u$, \be N_1=N+2u \hspace{1cm} N_2=N, \ee where $u$ will
be related to the difference in radii of the two spheres and from now
on we will restore $\alpha$ using dimensional analysis. Define \be
\eta^2_{\pm}&=&z^2\pm\frac{1}{2}\sqrt{\left(\frac{\alpha \mu}{6}\right)^4+16\alpha^2 v^2}\nonumber\\
\nu^2_{\pm}&=&z^2\pm \alpha v \ee and assuming that $u\geq 1$ (rather
than assuming $u\geq 0$ because the lower limit of the first
summation in eqn (\ref{eqn:VBeff}) has to be non-negative) we can
write the effective action so that $j$ always starts from $0$ and
finishes at $N-1$. \be \label{VeffSum} V_{eff}=-\frac{1}{\alpha}
\sum^{N-1}_{j=0}\biggl\{&&\hspace{-5mm}(2j+2u-1)\left[z^2+\left(\frac{\alpha
\mu}{3}\right)^2
(j+u)^2\right]^{\frac{1}{2}}\nonumber\\
     &&\hspace{-1cm}+ (2j+2u+3)\left[z^2+\left(\frac{\alpha \mu}{3}\right)^2(j+u+1)^2\right]^{\frac{1}{2}}\nonumber\\
     &&\hspace{-1cm}+ (2j+2u+1)\left[z^2+\left(\frac{\alpha \mu}{3}\right)^2(j+u)(j+u+1)\right]^{\frac{1}{2}}\nonumber\\
     &&\hspace{-1cm}+ 5(2j+2u+1)\left[z^2+\left(\frac{\alpha \mu}{3}\right)^2(j+u+1/2)^2\right]^{\frac{1}{2}}\nonumber\\
     &&\hspace{-1cm}+ (2j+2u+1)\Bigg(\left[\eta^2_{+}+\left(\frac{\alpha \mu}{3}\right)^2((j+u)(j+u+1)+\frac{1}{8})\right]^{\frac{1}{2}}\nonumber\\
    &&\hspace{3cm}+\left[\eta^2_{-}+\left(\frac{\alpha \mu}{3}\right)^2((j+u)(j+u+1)+\frac{1}{8})\right]^{\frac{1}{2}}\Bigg)\nonumber\\
     &&\hspace{-1cm}- 2(2j+2u)\left(\left[\nu^2_{+}+\left(\frac{\alpha \mu}{3}\right)^2(j+u+1/4)^2 \right]^{\frac{1}{2}}
        +\left[\nu^2_{-}+\left(\frac{\alpha \mu}{3}\right)^2(j+u+1/4)^2\right]^{\frac{1}{2}}\right)\nonumber\\
     &&\hspace{-1cm}- 2(2j+2u+2)\left(\left[\nu^2_{+}+\left(\frac{\alpha \mu}{3}\right)^2(j+u+3/4)^2\right]^{\frac{1}{2}}+\left[\nu^2_{-}
        +\left(\frac{\alpha \mu}{3}\right)^2(j+u+3/4)^2\right]^{\frac{1}{2}}\right)\nonumber\\
     &&\hspace{-1cm}- 2(2j+2u+1)\left[z^2+\left(\frac{\alpha \mu}{3}\right)^2(j+u)(j+u+1)\right]^{\frac{1}{2}}
     \biggr\}\nonumber\\
\ee

Let us write the above summation as \BE
V_{eff}=\sum_{j=0}^{N-1}{\cal V}(j)\EE and use Euler-Maclaurin sum
formula to convert it into integrals

\begin{eqnarray}\label{eqn:nuy}
V_{eff} = \int^{N}_0
 {\cal V}(j) dj-\frac{1}{2}[{\cal V}(N)+{\cal V}(0)]+\frac{1}{12}[{\cal V}'(N)-{\cal V}'(0)]\nonumber\\
\hspace{5cm} -\frac{1}{720}[{\cal V}'''(N)-{\cal V}'''(0)]+\cdots
\end{eqnarray}

It is useful to first see what we are  expecting from such an
integral. From eqn(\ref{VeffSum}), we see that a typical term of
${\cal V}(j)$  is roughly of the form:
\begin{eqnarray}
{\cal V}(j)&=&  \frac{1}{\alpha} j \sqrt{z^2 + \alpha v +\alpha^2 \mu^2 j^2} \nonumber \\
&=&  \mu N^2 \frac{j}{N} \sqrt{(\frac{j}{N})^2 +
\frac{1}{N^2}((\frac{z}{\alpha \mu})^2 + \frac{v}{\alpha \mu^2}) }
\end{eqnarray}

Plugging the above form of ${\cal V}$ into eqn (\ref{eqn:nuy}) and
defining new variables $\gamma = j/N$ and $\zeta
=\sqrt{((\frac{z}{\alpha \mu})^2 + \frac{v}{\alpha \mu^2})}$ , not
keeping track of the exact coefficients, we have:
\begin{eqnarray}
V_{eff} &=& \mu N^3 \int_{0}^1 d\gamma \  \gamma \sqrt{\gamma^2 + (\frac{\zeta}{N})^2 }
+ \mu N^2 \left(\sqrt{1 + (\frac{\zeta}{N})^2 } \right)\nonumber \\
&& \quad \quad + \mu N \left(\partial_\gamma \left(\gamma
\sqrt{\gamma^2  + (\frac{\zeta}{N})^2 }
\right)\right){\Bigg\arrowvert}^1_0+
\frac{\mu}{N} \left(\partial_\gamma^3 \left(\gamma \sqrt{\gamma^2  + (\frac{\zeta}{N})^2 } \right)\right){\Bigg\arrowvert}^1_0+ \cdots \nonumber \\
&=& \mu N^3 \bigg\{ F_0(\frac{\zeta}{N})+
\frac{1}{N}F_1(\frac{\zeta}{N})+ \frac{1}{N^2}F_2(\frac{\zeta}{N})
+\cdots \bigg\}
\end{eqnarray}
where $F_n$ are functions of $\frac{\zeta}{N} =\frac{1}{N}
\sqrt{((\frac{z}{\alpha \mu})^2 + \frac{v}{\alpha \mu^2})}$
originating from the $n$-th derivative of $\gamma$ (we shall see in
the next paragraph why we use $\frac{\zeta}{N}$ as the argument for
$F_n$ ). Note that each $F_n$ is weighted by a factor
$\frac{1}{N^n}$.

First we look at the term $F_0$.  Assuming in an power expansion of
$F_0(x)$ there exists a term $x^3$, after expanding in $\alpha$ it
would contribute to $V_{eff}$ a $v^4$ term of the form $\mu N^3
(\frac{\zeta}{N})^3 \sim \frac{\alpha v^4}{\mu^2 z^5}$, which has
the correct form of membrane interactions and has the correct order
of $N$ (see section \ref{MTSugraIntro}). If in the power expansion
of $F_0(x)$ there also exists a term $x^1$, then it would contribute
to $V_{eff}$ a $v^4$ term of the form $\mu N^3 (\frac{\zeta}{N})
\sim \frac{N^2 \alpha^3 v^4}{ z^7}$. This is the same expression as
graviton interactions. The details of the coefficients and whether
such terms exist in the power expansion of course has to be seen by
actually performing the integration, however as we will see later,
the integrals do produce terms of the correct interactions, in the
both the membrane and the graviton limit.

Next we look at $F_n$ for $n>0$.  The whole argument in the last
paragraph goes through, except now every term is weighted by an
extra factor of $\frac{1}{N^n}$. For example, the membrane like
interaction produced by $F_n$ looks like $ \frac{1}{N^n}
\frac{\alpha v^4}{\mu^2 z^5}$. This factor of $\frac{1}{N^n}$ means
that this term is in fact a matrix theory correction to
supergravity, because it vanishes as $N \rightarrow \infty$.
Therefore, we could see that in converting the summation into a
series of integrals, only the first one is needed for comparison
with supergravity. All the other $F_n$ with $n>0$ produces only
matrix theory corrections which is not the subject of interest here.

Now we go back to eqn (\ref{VeffSum}). As discussed above, we only
need the integral part $F_0$, which we calculate using Mathematica.
After calculating the integrals, $u$ is replaced by  the
supergravity variable $w=\frac{\alpha \mu}{3}u$, and the answer is
expanded first in large $N$, keeping only the leading order (which
is order $N^0$), and then expanded in small $\alpha$, keeping only
the $(\alpha)^1$ order (which according to Subsection
\ref{subsection:form_of_V} is the appropriate order to be compared
to supergravity in the membrane limit). Finally the answer is
converted back into Minkowski signature by sending
$v^2\rightarrow-v^2$. We obtain the following one-loop potential in
the membrane limit\footnote{This is only the effective potential in
the membrane limit because when we expand in large $N$ and keeping
only the lowest order we essentially send the radius of the spheres $r_0 \sim \alpha \mu N$
to infinity, thus compare with $z$ we have $\frac{z}{r_0} \ll 1$. Note also that the order the limits are taken is important. If the small $\alpha$ limit is taken before the large $N$ limit, implicitly we would be assuming $\frac{z}{\alpha \mu} \gg N$ which is the graviton limit. }:
\be V_{eff}=  \alpha
\left(\frac{(36v^2-z^2w^2)(108v^2-(4w^2+7z^2)\mu^2)}{1152
(w^2+z^2)^{5/2} \mu^2}\right). \ee

We find exact agreement when this expression is compared with the
supergravity light cone Lagrangian given in eqn
(\ref{eqn:deltaLlc_result}).

\section{Interpolation of the Effective Potential} \label{Interpolation}

\subsection{The Interpolation}

The purpose of this section is to find the effective potential that
interpolates between the membrane limit and the graviton limit. Due
to the complexity of the field equations on the supergravity side,
this problem could only be attacked on the matrix theory side. On
the matrix theory side, however, there is a subtlety that needs to
be taken into account before such a potential could be found. Here
we will analyse only the $v^4$ term, the $\mu^2 v^2$ term as well as
$\mu^4$ term can be studied in exactly the same way.

From the supergravity side, what we wish to find is more or less
clear. Near the membrane, when $\frac{z}{r_0} \ll 1$, we expect an
expansion like:
\begin{eqnarray}
V_{eff} = \frac{\alpha v^4}{\mu^2 z^5} (1+ \frac{z}{r_0} + ( \frac{z}{r_0})^2+ ( \frac{z}{r_0})^3+ ( \frac{z}{r_0})^4+ \cdots)
\end{eqnarray}

Far away from the membrane,  when $\frac{z}{r_0} \gg 1$, we expect an expansion:
\begin{eqnarray}
V_{eff} = \frac{\alpha v^4}{\mu^2 z^5} \frac{r_0^2}{z^2}(1+\frac{r_0}{z} + ( \frac{r_0}{z})^2+ ( \frac{r_0}{z})^3+ (\frac{r_0}{z})^4+ \cdots)
\end{eqnarray}
We have used $\frac{N^2 \alpha^3 v^4}{z^7} = \frac{\alpha v^4}{\mu^2 z^5} \frac{r_0^2}{z^2}$ to rewrite the graviton result so that it looks more similar to the membrane effective potential.

Therefore, we are basically looking for a function ${\cal C}(x)$
which appears in the effective potential in the following way:
\begin{eqnarray}
V_{eff} =  \frac{\alpha v^4}{\mu^2 z^5} {\cal C}(\frac{z}{r_0})
\end{eqnarray}
As one could see, both the graviton and the membrane action is in
this form. ${\cal C}(x)$ should have the appropriate limit at $x
\rightarrow 0$ and $x \rightarrow \infty$ to give the correct
potential at the membrane and the graviton limit respectively.
Physically it represents curvature corrections due to the finite size
of the spherical membrane.

So now we go to the matrix theory side to try to find ${\cal C}(x)$.
The subtlety is that the effective potential on the matrix theory
side not only includes curvature corrections but also the matrix
theory corrections to supergravity which we are not interested in,
not to mention that the sum over mass formula is incapable of
deducing such matrix theory corrections exactly \cite{Lee:2003kf}.

For our purpose of comparing with supergravity, therefore, all
matrix theory corrections to supergravity should be thrown away.
Such corrections appear on the matrix theory side as $1/N$
corrections, mixed together with the curvature corrections, and it
looks something like:
\begin{eqnarray} \label{CurvatureCorr}
V_{eff} =  \frac{\alpha v^4}{\mu^2 z^5}\left( {\cal C}_0(\frac{z}{r_0}) + \frac{1}{N} {\cal C}_1(\frac{z}{r_0}) + \frac{1}{N^2} {\cal C}_2(\frac{z}{r_0})+\cdots \right)
\end{eqnarray}
In such an expansion, only the ${\cal C}_0$ term should be kept. The readers are cautioned that naively sending $N$ to infinity will not give us the correct interpolating potential because $r_0$ also depends on $N$ and such limit would only result in the effective potential in the membrane limit.

There are many ways such matrix corrections could appear. For
example, in a typical matrix theory computation we may get terms of
the form:
\begin{eqnarray}
V_{eff} \sim  \frac{\alpha v^4}{\mu^2 z^7} ( z^2 + \alpha^2 \mu^2)
\end{eqnarray}
Rewriting the above gives:
\begin{eqnarray}
V_{eff} &\sim&  \frac{\alpha v^4}{\mu^2 z^5} ( 1 + \frac{\alpha^2 \mu^2 N^2}{z^2}\frac{1}{N^2}) \\
&\sim&  \frac{\alpha v^4}{\mu^2 z^5} ( 1 + \frac{r_0^2}{z^2}\frac{1}{N^2})
\end{eqnarray}
The second term could now be identified as a matrix theory correction and is irrelevant to us.

To isolate the curvature corrections (which we want) from the matrix
theory corrections (which we do not want), we look at
eqn(\ref{CurvatureCorr}) more carefully. We could see that since
$\frac{1}{N} = \frac{\alpha \mu}{6 r_0}$, all the matrix theory
corrections will appear in higher order in $\alpha$. Therefore to
get the interpolating effective potential from the matrix theory
side, we could follow the steps below:
\begin{enumerate}
\item Change the summation over $j$ in eqn(\ref{VeffSum}) into an
integral over $j$ from $0$ to $N$;
\item Replace $N$ by $ \frac{6 r_0}{\alpha \mu}$, and $u$ by $\frac{3w}{\alpha\mu}$;
\item Expand in small $\alpha$ and keeping only the lowest order (which shall turn out to be order
$\alpha^1$, with all lower orders vanishing ). This is the
interpolating effective potential.
\end{enumerate}

With Mathematica, the interpolating effective potential for two
spheres of the same radius ($w=0$) in Minkowski signature is found
to be:
\begin{multline} \label{VInt}
V_{eff} = \frac{\alpha}{\mu^2  z^5} \frac{36 v^2 -  z^2 \mu^2}{1152(4r_0^2 + z^2)^{5/2}}\\
\times \Bigg\{108 v^2 \Big(-z^5 +16 r_0^4 \sqrt{4r_0^2 + z^2} + 8 r_0^2 z^2\sqrt{4r_0^2 + z^2} + z^4\sqrt{4r_0^2 + z^2}\Big) \\
- z^2\mu^2 \Big(112 r_0^4\sqrt{4r_0^2 + z^2} + 7z^4 (-z +
\sqrt{4r_0^2 +z^2}) + 8 r_0^2 z^2 (-2 z + 7 \sqrt{4r_0^2 +
z^2})\Big)\Bigg\}
\end{multline}

We see that $V_{eff}$ always carries a factor of $36 v^2 -  z^2
\mu^2$, meaning the effective potential vanishes whenever
$v=\frac{\mu z}{6}$. This is expected because such configurations
correspond to circular orbits which preserve half of the
supersymmetries.

Expanding this potential in the membrane limit of large $r_0$ we get:
\begin{eqnarray}
V_{eff} = \frac{\alpha(3888 v^4 - 360 v^2 z^2 \mu^2 + 7 z^4 \mu^4)}{1152 \mu^2  z^5}
= \frac{\alpha(36 v^2 - z^2 \mu^2)(108 v^2 - 7 z^2 \mu^2)}{1152 \mu^2  z^5}
\end{eqnarray}
This result is of course identical to the matrix theory result (with
$w=0$) in section \ref{gaugeside} where the membrane limit was taken
in advance.

In the graviton limit of small $r_0$, after replacing $r_0$ by $\alpha \mu N/6$, we have:
\begin{eqnarray}
V_{eff} = \frac{N^2 \alpha^3 (720 v^4 - 56 v^2 z^2 \mu^2 + z^4 \mu^4)}{768 z^7}
= \frac{N^2 \alpha^3 (20 v^2 - z^2 \mu^2)(36 v^2 - z^2 \mu^2)}{768 z^7}
\end{eqnarray}

The two limits of the effective action could then be compared with
that of the supergravity side. Indeed from the expressions
(\ref{eqn:deltaLcmembrane}) and (\ref{eqn:deltalcgraviton}) we see
that we have perfect agreement.

In above we have only given the expression of the interpolating
potential for two membranes with the same radius ($w=0$). We have
also found the interpolating potential with $w$ included, using the
same steps given above. However we choose to omit the rather lengthy
expression here for brevity.

\subsection{Comparison with Shin and Yoshida}\label{subsection:comparison_with_shin}
As mentioned in section 5, ref. \cite{Shin:2003np}, considered the case of supersymmetric circular motion with angular frequency 
$\frac{\mu}{6}$ and found a flat
potential, which agrees with what we have found (see the comment under
eqn(\ref{VInt}) ).

In a subsequent paper, \cite{Shin:2004az}, the authors considered
the case of a slightly elliptical orbit with separation, $z$, \be
z&=&\sqrt{(r_2+\epsilon)\cos^2\left(\frac{\mu t}{6}\right)+(r_2-\epsilon)\sin^2\left(\frac{\mu t}{6}\right)}\nonumber\\
 &=&\sqrt{r_2^2+\epsilon^2+2r_2\epsilon \cos\left(\frac{\mu t}{3}\right)}
\ee and velocity, $v$, \be
v&=&\frac{\mu}{6}\sqrt{\left((r_2+\epsilon)\sin^2(\frac{\mu t}{6})+(r_2-\epsilon)\cos^2\left(\frac{\mu t}{6}\right)\right)}\nonumber\\
 &=&\frac{\mu}{6}\sqrt{\left((r_2^2+\epsilon^2-2r_2\epsilon \cos\left(\frac{\mu t}{3}\right)\right)}.
\ee where $\epsilon$ is the small expansion parameter for the
eccentricity of the orbit. They considered the large separation
limit, $r_2 \gg 0$, and found an effective action, eqn(1.2) of
\cite{Shin:2004az}, \be\label{eqn:shinandyoshida}
\Gamma_{eff}&=&\epsilon^4\int dt \hspace{1mm}(\alpha^3\mu^4N^2
)\left(\frac{35}{2^7\cdot3}\frac{1}{r_2^7}-\frac{385(\alpha\mu
N)^2}{2^{11}\cdot3^3}
(4-\frac{1}{N^2})\frac{1}{r_2^9}\right)\nonumber\\
&=&\frac{35}{384}\epsilon^4\int dt \hspace{1mm}(\alpha^3\mu^4N^2
)\left(\frac{1}{r_2^7}-\frac{11(\alpha\mu
N)^2}{36}(1-\frac{1}{4N^2})\frac{1}{r_2^9}\right) \ee after
expanding to $O(\epsilon^4)$ and $O(1/r_2^9)$. Note that in the
equation above we have restored $\mu$ and $\alpha$, and set
$N_1=N_2=N$, $r_1=0$. To compare this with our result we substitute
the above expressions of $z(t)$ and $v(t)$ into our interpolating
potential eqn (\ref{VInt}) and expand in the parameters $1/r_2$ and
$\epsilon$. As we are only comparing effective actions we average
our potential over one period of oscillation. We find for our
time-averaged potential, \be
V_{eff}&=&\alpha\epsilon^4\mu^2\frac{105}{32}r_0^2\left(\frac{1}{r_2^7}-11\frac{r_0^2}{r_2^9}\right)\nonumber\\
&=&\frac{35}{384}\alpha^3\epsilon^4\mu^4N^2\left(\frac{1}{r_2^7}-\frac{11(\alpha\mu
N)^2}{36}\frac{1}{r_2^9}\right) \ee (where to reach the final line
we used $r_0=\frac{\alpha\mu N}{6}$) which agrees with eqn
(\ref{eqn:shinandyoshida}) after throwing away the $\frac{1}{N^2}$
matrix theory corrections in the latter. In calculating the matrix theory effective potential in section 
\ref{gaugeside} we assumed a constant velocity. However, as we can see by this comparison, as long as we 
ignore matrix theory corrections, it  leads to the correct result. Thus we see that we can consistently 
neglect acceleration terms  in the effective potential as discussed earlier.

\section{Discussion and Future Directions} \label{sec:discussion}
One immediate generalization of the work reported in this paper is
to consider more complicated probe configurations (recall that in
this paper we have restricted our attention to a probe membrane
which is spherical and has no velocity in the $x^1$ through $x^3$
directions). For example, we could consider deforming the probe
membrane so that it is no longer a perfect sphere. It means that the
coordinates of the membrane $X^A$ will now be some general functions
of $\theta$ and $\phi$, and in particular it no longer has to appear
only as a point in the $4$ through $9$ directions. We can also give
the probe membrane a nonzero velocity in the $x^1$ through $x^3$
direction. These generalizations give more interesting dynamics and
can be fairly easily carried out. On the supergravity side, this
just requires putting the relevant probe configuration into the
light-cone Lagrangian; on the matrix theory side, this requires
replacing the background configurations $B^A_{(2)}$'s used in this
paper with some more general configurations.

In this paper we take the source to be a static spherical membrane
preserving all of the sixteen linearly realized supersymmetries. As
we know, there are other interesting membrane configurations that
are permitted by the pp-wave background, for example, the static
hyperbolic brane which preserves eight supersymmetries, given by
\cite{Bak:2002rq}. It would be interesting to investigate
gauge/gravity duality in the case where these static objects are
sources and interact with some graviton or membrane probe. On the
supergravity side, we can apply our general formalism of
diagonalizing the field equations as well as the near-membrane
expansion to find the metric and three-form perturbations produced
by these sources. On the matrix theory side, for the aforementioned
hyperbolic membrane, one could use the unitary representations of
$SO(2,1)$ worked out in
\cite{Dixon:1989cg,Maldacena:2000hw,Morariu:2002tx} to expand the
fluctuating fields and compute the one-loop effective potential.

Moreover, it is evidently desirable to push our computation on the
supergravity side beyond the near-membrane expansion, finding the
solution to the field equations which will then give the
$\delta{\cal L}_{lc}$ to be compared with the fully interpolating
potential (\ref{VInt}) found on the matrix theory side.

The membrane-interaction we described in this paper has no
longitudinal momentum transfer. However, as we can see, our
supergravity computation can be quite easily generalized to nonzero
longitudinal momentum transfer, by not setting $k_-$ to zero when
solving the field equations (e.g., see the comments made after eqn.
(\ref{eqnLlc0})). On the gauge theory side, this requires an
instanton compuation. We hope to say more on this in the near
future.

There are also M-theory pp-wave backgrounds with less supersymmetries, and
matrix theories in these pp-waves backgrounds have been proposed
\cite{Cvetic:2002si, Iizuka:2002ra}. It would be interesting to
investigate the gauge/gravity duality in these less supersymmetric
settings. Quite recently a matrix theory dual of strings on the
maximally supersymmetric ten-dimensional IIB pp-wave background was
conjectured by \cite{Sheikh-Jabbari:2004ik}. This matrix theory is
the regularization of the action of D3-branes (tiny gravitons). It
would be very interesting to investigate this conjecture along the lines
of this paper and \cite{Lee:2003kf}.

\section*{Acknowledgments}
The authors would like to thank John Schwarz for helpful advice and support
during the summer. We would also like to thank I. Swanson for useful discussions. This work was 
supported by DOE grant DE-FG03-92ER40701. X.Wu is also supported in part by DOE grant DE-FG01-00ER45832 and NSF grant PHY/0244811.



\begin{thebibliography}{99}


\bibitem{Berenstein:2002jq}
D.~Berenstein, J.~M.~Maldacena and H.~Nastase, ``Strings in flat
space and pp waves from N = 4 super Yang Mills,'' JHEP {\bf 0204},
013 (2002) [arXiv:hep-th/0202021].

\bibitem{Lee:2003kf}
H.~K.~Lee and X.~k.~Wu, ``Two-graviton interaction in pp-wave
background in matrix theory and supergravity,'' Nucl.\ Phys.\ B {\bf
665}, 153 (2003) [arXiv:hep-th/0301246].

\bibitem{BB1}
K.~Becker and M.~Becker,
``A two-loop test of M(atrix) theory,''
Nucl.\ Phys.\ B {\bf 506}, 48 (1997)
[arXiv:hep-th/9705091].

\bibitem{BB2}
K.~Becker and M.~Becker,
``On graviton scattering amplitudes in M-theory,''
Phys.\ Rev.\ D {\bf 57}, 6464 (1998)
[arXiv:hep-th/9712238].

\bibitem{BB3}
K.~Becker and M.~Becker,
``Complete solution for M(atrix) theory at two loops,''
JHEP {\bf 9809}, 019 (1998)
[arXiv:hep-th/9807182].

\bibitem{BBPT}
K.~Becker, M.~Becker, J.~Polchinski and A.~A.~Tseytlin,
``Higher order graviton scattering in M(atrix) theory,''
Phys.\ Rev.\ D {\bf 56}, 3174 (1997)
[arXiv:hep-th/9706072].

\bibitem{OY1}
Y.~Okawa and T.~Yoneya,
``Multi-body interactions of D-particles in supergravity and matrix  theory,''
Nucl.\ Phys.\ B {\bf 538}, 67 (1999)
[arXiv:hep-th/9806108].

\bibitem{OY2}
Y.~Okawa and T.~Yoneya,
``Equations of motion and Galilei invariance in D-particle dynamics,''
Nucl.\ Phys.\ B {\bf 541}, 163 (1999)
[arXiv:hep-th/9808188].

\bibitem{Okawa}
Y.~Okawa,
``Higher-derivative terms in one-loop effective action for general  trajectories of D-particles in matrix theory,''
Nucl.\ Phys.\ B {\bf 552}, 447 (1999)
[arXiv:hep-th/9903025].
   
\bibitem{Shin:2003np}
H.~Shin and K.~Yoshida, ``One-loop flatness of membrane fuzzy sphere
interaction in plane-wave matrix model,'' Nucl.\ Phys.\ B {\bf 679},
99 (2004) [arXiv:hep-th/0309258].

\bibitem{Shin:2004az}
H.~Shin and K.~Yoshida,
``Membrane fuzzy sphere dynamics in plane-wave matrix model,''
arXiv:hep-th/0409045.

\bibitem{Park:2002cb}
J.~H.~Park,
``Supersymmetric objects in the M-theory on a pp-wave,''
JHEP {\bf 0210}, 032 (2002)
[arXiv:hep-th/0208161].

\bibitem{Becker:1997xw}
K.~Becker, M.~Becker, J.~Polchinski and A.~A.~Tseytlin,
``Higher order graviton scattering in M(atrix) theory,''
Phys.\ Rev.\ D {\bf 56}, 3174 (1997)
[arXiv:hep-th/9706072].

\bibitem{xinkaithesis}
Xinkai Wu, Ph.D. thesis, available at
http://etd.caltech.edu/etd/available/etd-05252004-230238/

\bibitem{Kimura:2003um}
T.~Kimura and K.~Yoshida, ``Spectrum of eleven-dimensional
supergravity on a pp-wave background,'' {\it Phys.\ Rev.\ D} {\bf
68}, 125007 (2003) [arXiv:hep-th/0307193].

\bibitem{Polchinski:1997pz}
J.~Polchinski and P.~Pouliot, ``Membrane scattering with M-momentum
transfer,'' Phys.\ Rev.\ D {\bf 56}, 6601 (1997)
[arXiv:hep-th/9704029].


\bibitem{Dasgupta:2002hx}
K.~Dasgupta, M.~M.~Sheikh-Jabbari and M.~Van Raamsdonk, ``Matrix
perturbation theory for M-theory on a PP-wave,'' JHEP {\bf 0205},
056 (2002) [arXiv:hep-th/0205185].


\bibitem{Das:2003yq}
S.~R.~Das, J.~Michelson and A.~D.~Shapere, ``Fuzzy spheres in
pp-wave matrix string theory,'' Phys.\ Rev.\ D {\bf 70}, 026004
(2004) [arXiv:hep-th/0306270].


\bibitem{Bak:2002rq}
D.~s.~Bak, ``Supersymmetric branes in PP wave background,'' Phys.\
Rev.\ D {\bf 67}, 045017 (2003) [arXiv:hep-th/0204033].


\bibitem{Dixon:1989cg}
L.~J.~Dixon, M.~E.~Peskin and J.~Lykken, ``N=2 Superconformal
Symmetry And SO(2,1) Current Algebra,'' Nucl.\ Phys.\ B {\bf 325},
329 (1989).


\bibitem{Maldacena:2000hw}
J.~M.~Maldacena and H.~Ooguri, ``Strings in AdS(3) and SL(2,R) WZW
model. I,'' J.\ Math.\ Phys.\  {\bf 42}, 2929 (2001)
[arXiv:hep-th/0001053].

\bibitem{Morariu:2002tx}
B.~Morariu and A.~P.~Polychronakos, ``Quantum mechanics on
noncommutative Riemann surfaces,'' Nucl.\ Phys.\ B {\bf 634}, 326
(2002) [arXiv:hep-th/0201070].


\bibitem{Cvetic:2002si}
M.~Cvetic, H.~Lu and C.~N.~Pope, ``M-theory pp-waves, Penrose limits
and supernumerary supersymmetries,'' Nucl.\ Phys.\ B {\bf 644}, 65
(2002) [arXiv:hep-th/0203229].


\bibitem{Iizuka:2002ra}
N.~Iizuka, ``Supergravity, supermembrane and M(atrix) model on
pp-waves,'' Phys.\ Rev.\ D {\bf 68}, 126002 (2003)
[arXiv:hep-th/0211138].


\bibitem{Sheikh-Jabbari:2004ik}
M.~M.~Sheikh-Jabbari, ``Tiny graviton matrix theory: DLCQ of IIB
plane-wave string theory, a conjecture,'' arXiv:hep-th/0406214.





\end{thebibliography}
\end{document}